\documentclass[amsmath, amsthm, amssymb, floatfix, reprint, twocolumn, notitlepage, nofootinbib, 10pt]{revtex4-1}

\usepackage{amsthm}
\makeatletter
\newtheorem*{rep@theorem}{\rep@title}
\newcommand{\newreptheorem}[2]{%
\newenvironment{rep#1}[1]{%
 \def\rep@title{#2 \ref*{##1}, repeated}%
 \begin{rep@theorem}}%
 {\end{rep@theorem}}}
\makeatother

\newtheorem{theorem}{Theorem}
\newreptheorem{theorem}{Theorem}
\newtheorem{lemma}{Lemma}
\newreptheorem{lemma}{Lemma}
\newtheorem{problem}{Problem}
\newreptheorem{problem}{Problem}

\newreptheorem{assumption}{Assumption}
\newtheorem{definition}{Definition}
\newreptheorem{definition}{Definition}
\newtheorem{proposition}{Proposition}

\usepackage[utf8]{inputenc}
\usepackage{microtype,bm,bbm,graphicx,booktabs,times}
\usepackage[usenames,dvipsnames]{xcolor}
\usepackage{setspace}
\selectfont{}

\usepackage{hyperref}
\usepackage{braket}
\usepackage{subfiles} 
\usepackage{upgreek}
\usepackage{color}
\hypersetup{colorlinks,allcolors=black,linktoc=all}
\usepackage[capitalize]{cleveref}
\crefformat{section}{Sec.~#2#1#3}

\newcommand{\edit}[1]{{\color{black} #1}}

\newcommand{\norm}[1]{\left\Vert{#1}\right\Vert}

\DeclareMathOperator*{\argmin}{arg\,min}
\newcommand{\abs}[1]{\left | {#1}\right |}

\begin{document}
\title{Simulability transitions in continuous-time dynamics of local open quantum systems}
\author{Rahul Trivedi$^{1, 2}$}
\email{rahul.trivedi@mpq.mpg.de}
\author{J.~Ignacio Cirac$^{1, 2}$}
\address{$^1$Max-Planck-Institut für Quantenoptik, Hans-Kopfermann-Str.~1, 85748 Garching, Germany.\\
$^2$Munich Center for Quantum Science and Technology (MCQST), Schellingstr. 4, D-80799 Munich, Germany.}

\date{\today}

\begin{abstract}
{We analyze the complexity of classically simulating continuous-time dynamics of locally interacting quantum spin systems with a constant rate of entanglement breaking noise. We prove that a polynomial time classical algorithm can be used to sample from the  {state of the spins when the rate of noise is higher than a threshold determined by the strength of the local interactions}. Furthermore, by encoding a 1D fault tolerant quantum computation into the dynamics of spin systems arranged on two or higher dimensional grids, we show that for  {several noise channels},  {the problem of weakly simulating the output state of both purely Hamiltonian and purely dissipative dynamics is expected to be hard in the low-noise regime.}}
\end{abstract}
\maketitle

\emph{Introduction}. {Locally interacting spin systems are of fundamental interest in many-body physics, and also describe engineered quantum systems that underly quantum information technologies.} Consequently, there is great interest in developing classical algorithms for simulating their dynamics \cite{vidal2004efficient, vidal2003efficient, muller2012tensor, rams2020breaking, zaletel2015time}. It is recognized that simulating quantum spin system dynamics on classical computers is generically hard since they can encode quantum computations \cite{watrous1995one, vollbrecht2006reversible, schuch2008entropy}. However, strong interaction with an external environment prevents significant entanglement of the individual spins \cite{werner2016positive, del2018tensor, jaschke2018one, luchnikov2019simulation, mc2021stable, nakano2021tensor}.  Physical intuition suggests that a simulability transition occurs on tuning the strength of the system-environment interaction i.e.~the spin system transitions from a \edit{classically tractable} {phase, whose dynamics can be simulated on a classical computer in time that scales at most polynomially with the number of spins, to a classically intractable phase}.

This expectation has been made rigorous in the context of \emph{circuit model} (or discrete-time model) of noisy quantum computation \cite{aharonov2000quantum, aharonov1996polynomial, aharonov2008fault, gottesman2000fault, benquantum, virmani2005classical, harrow2003robustness}. It was shown very early on that a simulability transition is expected for the circuit model of quantum computation on tuning the rate of noise.  {For sufficiently high rate of noise,  provably efficient classical algorithms to simulate quantum circuits \cite{aharonov2000quantum, aharonov1996polynomial} have been provided}. Moreover, the threshold theorem for quantum computation \cite{aharonov2008fault, gottesman2000fault} implied that if the noise is below a certain threshold and fresh auxillary qubits are available, then a {quantum computation can be encoded into a noisy quantum circuit.}   {The requirement of fresh auxillary qubits was subsequently relaxed for quantum circuits in two or higher dimensions when the noise was not depolarizing \cite{benquantum}}.

Less attention has been paid to simulability transitions in \emph{continuous-time dynamics}. Not only does it underly the discrete-time circuit model, it is also physically more relevant for analyzing near term analogue quantum simulators \cite{blatt2012quantum, schindler2013quantum, kim2010quantum, nguyen2018towards, schafer2020tools, houck2012chip, raftery2014observation, fitzpatrick2017observation}. While some studies have focussed on simulability transitions in bosonic systems as a function of evolution time \cite{maskara2019complexity, muraleedharan2019quantum, deshpande2018dynamical}, theoretical results on the simulability transitions as a function of noise strength have thus far only been provided for fermionic systems  \cite{shtanko2021complexity}.

{In this paper, we study simulability transition with noise rate in spatially locally interacting spin-systems. We consider an open system of $n-$spins arranged on a $d-$dimensional lattice ($\mathbb{Z}^d$) and initially in a product state. \edit{Within the Born-Markov approximation \cite{davies1974markovian}}, the state of the spins $\rho(t)$ is governed by a quantum Lindblad equation $d \rho(t)/dt  = \mathcal{L}(t) \rho(t)$, with the Lindbladian
\begin{align}\label{eq:general_lindbladian}
&\mathcal{L}(t) = -i[H(t), \rho(t)] +\nonumber\\
&\qquad \sum_{\alpha} \bigg(L_\alpha(t) \rho(t) L_\alpha^\dagger(t) - \frac{1}{2}\{\rho(t), L_\alpha^\dagger(t) L_\alpha(t) \}\bigg),
\end{align}
where $H(t)$ is the (possibly time-dependent) Hamiltonian, and $L_\alpha(t)$ are (possibly time-dependent) jump operators. Here, we study a restricted class of master equations with generators $\mathcal{L}(t)$ of the form
\begin{align}\label{eq:master_equation_paper}
\mathcal{L}(t)= \mathcal{L}_0(t)  + \kappa \sum_{i = 1}^n \big(\mathcal{N}_i - \text{id}\big),
\end{align}
\edit{where `$\text{id}$' is the identity channel. The generator of this master equation has two terms --- the first term, $\mathcal{L}_0(t)$, is a Lindbladian (i.e.~of the form of Eq.~\ref{eq:general_lindbladian}) that models interactions between different spins and the second term captures noise, modelled by a channel $\mathcal{N}_i$ on the $i^\text{th}$ spin, acting at a constant rate $\kappa$. We do not restrict ourselves to $\mathcal{L}_0(t)$ being described by only a Hamiltonian since even dynamics described by Lindbladians with only jump operators (albeit acting simultaneously on multiple spins) can be classically intractable \cite{{verstraete2009quantum}}.  } 

We constrain $\mathcal{L}_0(t)$ to be geometrically local with interaction range $R$ and with a uniformly bounded interaction strength $J$ i.e.~$\mathcal{L}_0(t)$ permits a representation
\begin{align}\label{eq:lindbladian_local_terms}
\mathcal{L}_0(t) = \sum_{\Lambda \subset \mathbb{Z}^d} \mathcal{L}_0^\Lambda(t),
\end{align}
where $\mathcal{L}_0^\Lambda(t)$ is a Lindbladian which is identity on spins outside $\Lambda$ with $\text{diam}(\Lambda) \leq R$ and $\norm{\mathcal{L}_0^\Lambda(t)}_{1 \to 1} \leq J$.  The noise channel $\mathcal{N}_i$ is assumed to be entanglement breaking \cite{horodecki2003entanglement} --- examples of such noise channels could include local depolarizing noise ($\mathcal{N}_i(\rho) = \text{tr}_i(\rho) I / 2$), dephasing noise ($\mathcal{N}_i(\rho) = (\rho + Z_i \rho Z_i) / 2$) and amplitude dampling noise ($\mathcal{N}_i(\rho) = \text{tr}_i(\rho) \ket{0}\bra{0}$). Throughout this paper, we consider evolution times $t$ that scale at most as poly($n$).}

{In the high noise regime, we show that this problem is classically tractable.} {Our proof strategy, inspired by previous results for quantum-circuits \cite{aharonov2000quantum, skinner2019measurement}, is to identify a map between the quantum dynamics and a percolation problem. However, unlike the discrete-time setting, where the dynamics respect a strict light-cone thus making this mapping direct, the continuous-time dynamics for local Lindbladians only has an approximate light-cone \cite{osborne2006efficient}. Our key technical contribution is to show that an approximation of the continuous-time dynamics can be mapped to a \emph{correlated} percolation problem, which we prove percolates at a sufficiently high rate of noise.}

{We next consider the complementary low noise regime and study the worst-case hardness of this problem. The threshold theorem for quantum computation \cite{aharonov2008fault, gottesman2000fault} already suggests that local Lindbladians are classically intractable below a noise threshold. This is so because a local Lindbladian can be chosen to fault-tolerantly encode \emph{any} given quantum computation \cite{aharonov2008fault, gottesman2000fault}, which cannot be efficiently classically simulated. \edit{However, it is often of interest to study models where $\mathcal{L}_0(t)$ either captures Hamiltonian interactions i.e.~where $\mathcal{L}_0(t)\rho = -i[H(t), \rho]$ for some Hamiltonian $H(t)$ (e.g.~out of equilibrium many-body systems \cite{eisert2015quantum,kaufman2016quantum,d2016quantum}), or purely dissipative interactions i.e.~where $\mathcal{L}_0(t) \rho = \sum_k L_k(t) \rho L_k^\dagger(t) - \{L_k^\dagger(t) L_k(t), \rho\}/2$ for some jump operators $L_k(t)$ (e.g.~superradiance in many-body quantum optics \cite{masson2020many, sierra2022dicke, masson2022universality}). While both of these classes of systems are known to be hard to classically simulate when $\kappa = 0$ \cite{nagaj2008hamiltonian, vollbrecht2008quantum, verstraete2009quantum}, their worst-case hardness in the low (but non-zero) noise regime does not follow from a direct application of the threshold theorem.}

We show that, for both of these classes and for noise rates below a threshold, it is unlikely that an efficient classical algorithm can simulate Eq.~\ref{eq:master_equation_paper} in two or higher dimensions for arbitrary noise channels $\mathcal{N}_i$. More specifically, {by an adaptation of Ref.~\cite{benquantum} to continuous-time, we identify a class of non-unital noise channels (which includes, e.g.,~the amplitude damping channel) such that Eq.~\ref{eq:master_equation_paper} with purely Hamiltonian $\mathcal{L}_0(t)$ is classically intractable below a noise threshold. We then consider $\mathcal{L}_0(t)$ to be purely dissipative, which is classically intractable without noise \cite{verstraete2009quantum}. We show that, for amplitude damping or dephasing noise and when $\kappa$ is below a threshold, the dissipative dynamics can encode a postselected quantum computation, and hence is expected to be classically intractable.}

\emph{Results}. {Our first result considers the high-noise regime of Eq.~\ref{eq:master_equation_paper}, and shows its classical tractability.

\begin{theorem}
\label{theorem:theorem_1}
For $\kappa > \kappa_\textnormal{th}$, where $\kappa_\textnormal{th}$ depends on the lattice dimension $d$, interaction range $R$ and interaction strength $J$, there is  {a $\textnormal{poly}(n, 1/\varepsilon)$ randomized} classical polynomial-time algorithm to sample within $\varepsilon$ total variation distance of $\rho(t)$ obtained on evolving Eq.~\ref{eq:master_equation_paper} for $t$ scaling at most as $\textnormal{poly}(n)$.
\end{theorem}}

Our general strategy for the classical algorithm is to map Eq.~\ref{eq:master_equation_paper} to a percolation problem. This has previously been done for the local unitary circuits \cite{aharonov2000quantum, skinner2019measurement}, where at sufficiently high noise rate, the effective percolation problem is subcritical \cite{bazant2000largest} and the circuit can be \emph{exactly} broken into small non-interacting clusters which permit individual contraction. However, unlike local unitary circuits, the continuous-time dynamics does not respect an exact light-cone \cite{osborne2006efficient} and consequently this mapping is not direct. We circumvent this issue by mapping a trotterized approximation of this dynamics to a \emph{correlated percolation problem} that is then shown to percolate at sufficiently high noise rates $\kappa$. 

\begin{figure*}
\centering
\includegraphics[scale=0.42]{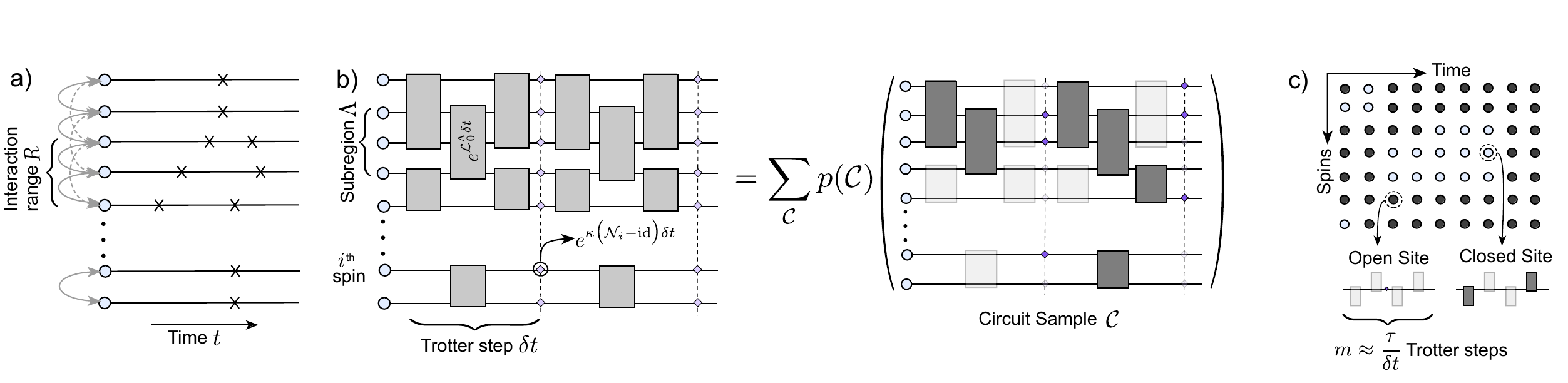}
\caption{{Schematic depiction of the mapping of a continuous time model, to an equivalent percolation problem. For simplicity, we only depict a 1D setting. (a) The continuous-time model, where the spins interact with each through a local Lindbladian and with the crosses depicting the entanglement breaking noise occuring at a rate $\kappa$. (b) Trotterization of the continuous-time evolution, with the grey rectangles representing the channels resulting from $\mathcal{L}_0(t)$ and the purple rhombi being the trotterized single-qubit noise channel. The trotterized channels are then sampled from, with the faded channels being sampled to identity. (c) blocking $m \approx \tau / \delta t$ time-steps together for each qubit and identifying this block of time-steps with a site on the percolating lattice. The site is declared open if the qubit associated with the site experiences entanglement breaking noise at least once and does not couple to any of the neighbouring qubits via the channels obtained on trotterization of the Lindbladian $\mathcal{L}_0(t)$.}}
\label{fig:fig_cont_time_perc}
\end{figure*}

 \emph{Proof sketch for theorem 1}. The steps in the proof are depicted in Fig.~\ref{fig:fig_cont_time_perc} --- we first trotterize the evolution with time-step $\delta t = O(\varepsilon / \text{poly}(n))$ chosen to incur a total variation error $\leq O(\varepsilon )$ [Fig.~\ref{fig:fig_cont_time_perc}(a) to (b)]. Next, we approximate the channels resulting from the trotterization of $\mathcal{L}_0(t)$, as a convex combination of identity, applied with probability $1 - O(J)\delta t$ and another channel, applied with probability $O(J)\delta t$. The probability distribution at the output of the trotterized circuit, $p(\textbf{x})$, is then expressed as 
\[
p(\textbf{x}) = \sum_{\mathcal{C}} p(\mathcal{C}) p(\textbf{x} | \mathcal{C}),
\]
where the summation is over circuits $\mathcal{C}$ [Fig.~\ref{fig:fig_cont_time_perc}(b)] obtained by randomly choosing between (i) identity or otherwise instead of the trotterization of local Lindbladian, and (ii) identity or $\mathcal{N}_i$ instead of the noise channel. $p(\mathcal{C})$ is the probability of choosing the circuit instance $\mathcal{C}$ thus obtained and $p(\textbf{x} | \mathcal{C})$ is the probability of obtaining $\textbf{x}$ at the output of $\mathcal{C}$.

Next, we use percolation theory to show that with high probability over $\mathcal{C}$, $p(\textbf{x} | \mathcal{C})$ can be efficiently sampled from on a classical computer. We first map sampling from $p(\mathcal{C})$ to a percolation problem on $\mathbb{Z}^{d + 1}$ [Fig.~\ref{fig:fig_cont_time_perc}(c)] --- a site in this equivalent percolation problem is associated with a qubit and a block of $m$ trotterized time steps, where $m \approx \tau / \delta t$ for some $\tau > 0$. For a sampled circuit $\mathcal{C}$, the site is declared open if the associated qubit experiences the noise channel at least once in the $m$ associated time-steps, \emph{and} all channels arising from the local Lindbladian acting on the qubit are replaced with identity, else it is declared closed. Note that this percolation problem is \emph{correlated} i.e. the state of each site is dependent on its neighbourhood. However, we show that for sufficiently large $\kappa$, $\tau$ can be chosen such that the percolation problem is subcritical. Similar to the discrete-time case~\cite{aharonov2000quantum}, the sizes of the clusters in the subcritically percolated lattice are almost surely $O(\text{log}\ n)$ \cite{bazant2000largest}, which allows us to classically compute $p(\textbf{x} | \mathcal{C})$ and its marginals in polynomial time, and thus draw a sample from the output of $\mathcal{C}$. A detailed proof is provided in the supplement. \hfill\(\square\)

 {Our next two results deal with the low-noise regime. Our first result considers local Hamiltonian dynamics, i.e.~$\mathcal{L}_0^\Lambda(t)$ in Eq.~\ref{eq:lindbladian_local_terms} satisfies $\mathcal{L}_0^\Lambda(t) \rho = -i[H^\Lambda(t), \rho]$ for some $H^\Lambda(t)$, and shows its low-noise insimulability. We restrict ourselves to entanglement breaking noise channels $\mathcal{N}_i$ of the form
\begin{align}\label{eq:form_N}
\mathcal{N}_i(\rho) = \text{Tr}_i(P \rho)\otimes \ket{\alpha}\bra{\alpha} + \text{Tr}_i(Q \rho) \otimes \ket{\beta}\bra{\beta},
\end{align}
where $\{P, Q\}$ is a single-qubit POVM, \edit{with $P  - I / 2$ being positive-definite} and $\{\ket{\alpha}, \ket{\beta}\}$ is an orthonormal basis for the qubit Hilbert space, which is then applied on the $i^\text{th}$ qubit. The channel $\mathcal{N}_i$ thus maps any initial state of the $i^\text{th}$ qubit to a mixture of $\ket{\alpha}, \ket{\beta}$ with a higher probability of being in $\ket{\alpha}$. An example of such a channel would be an amplitude damping channel.}  {The proof of this result is a straightforward adaption of the discrete-time fault-tolerant construction previously used in Ref.~\cite{benquantum}, the only additional ingredient needed being the analysis of how faults in the unitary gates (as opposed to before or after them) do not impact the threshold theorem.}
\begin{figure*}
\includegraphics[scale=0.47]{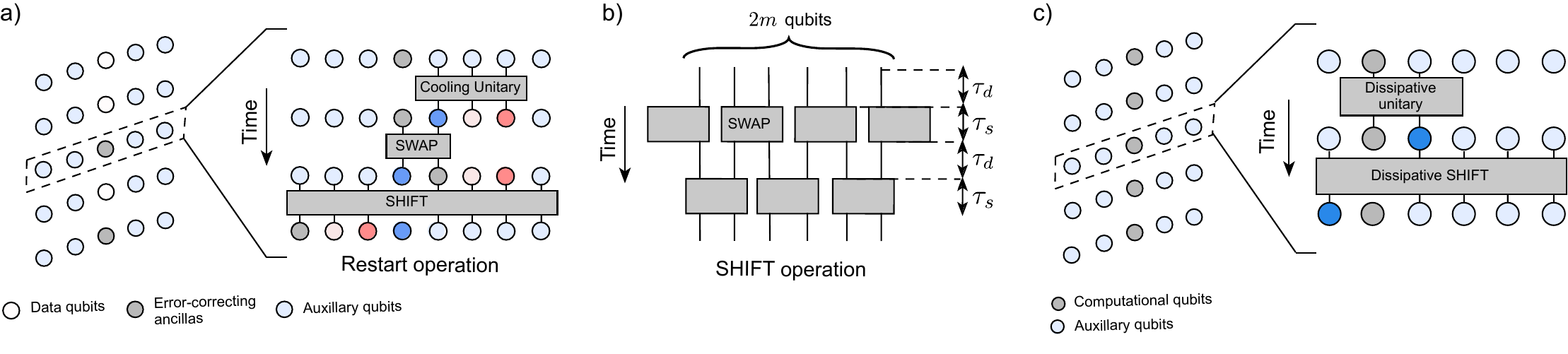}
\caption{{(a) Construction of worst-case example of Eq.~\ref{eq:master_equation_paper} when $\mathcal{L}_0(t)$ is a local Hamiltonian in 2D --- one column is used for encoding a 1D fault-tolerant quantum computation and RESTART operation is implemented using the auxillary qubits in the same row. The steps in the RESTART operation, also schematically depicted, include cooling the auxillary qubits, swap with ancilla and shift for the next restart operation. (b) The SHIFT operation implemented by layers of SWAP gates (which take time $\tau_s$) followed by allowing the noise to act on the individual qubits for time $\tau_d$. The SWAP gates are faulty due to the noise, but the subsequent time interval $\tau_d$ is used to drive the swapped qubits to the fixed point of the noise channel. (c) Construction of worst-case example of Eq.~\ref{eq:master_equation_paper} when $\mathcal{L}_0(t)$ is a local on a 2D lattice and purely dissipative. One column of qubits is again used to encode a 1D fault-tolerant quantum computation, with the remaining qubits used as the clock qubits to implement the involved unitaries dissipatively.}}
\label{fig:fig_2_4}
\end{figure*}

\begin{theorem}
\label{theorem:cont_time_circuit}
{If $\mathcal{N}_i$ is of the form described in Eq.~\ref{eq:form_N}, then for qubits arranged on two or higher dimensional lattices and for $\kappa$ below a threshold, there are instances of Eq.~\ref{eq:master_equation_paper} with $\mathcal{L}_0(t)$ being a local Hamiltonian that cannot be weakly simulated on a classical computer within a small total variation error\footnote{A family of $n-$qubit quantum circuits is said to be weakly simulable within $\varepsilon$-total variation error if a classical computer can be used to sample in $\textnormal{poly}(n)$ time within a probability distribution $p_\textnormal{cl}$ such that $\norm{p - p_\textnormal{cl}}_1 \leq \varepsilon$, where $p$ is the probability distribution at the output of the quantum circuit.} unless $\textnormal{BQP = BPP}$.}
\end{theorem}
 {\emph{Proof sketch for theorem 2}. We will restrict ourselves to two-dimensional lattices. We first briefly review the discrete-time construction of Ref.~\cite{benquantum} --- the key idea is to fault-tolerantly encode 1D local quantum circuits, which can perform arbitrary quantum computations \cite{watrous1995one, vollbrecht2006reversible}, in the continuous-time model. It has been previously established that fault tolerance can be achieved with just nearest neighbour unitary gates in 1D \cite{aharonov2008fault} if a RESTART operation (i.e.~a quantum channel which replaces a qubit with a known pure state, say $\ket{0}$) is accessible. Ref.~\cite{benquantum} proposed to exploit the noise channel to implement the RESTART gate. Given a 2D grid of qubits [Fig.~\ref{fig:fig_2_4}(a)], qubits in one column of the lattice are used as the computational qubits, comprising of data qubits (on which the quantum computation is performed) and ancilla qubits (which are used to perform error correction and need to be restarted). To restart an ancilla qubit, they utilize the qubits, henceforth called the auxillary qubits, in the row containing the ancilla. These qubits are initialized in the fixed point of $\mathcal{N}_i$ (and hence remain in it at all times), and when the ancillas need to be RESTARTED, a (constant) number of auxillary qubits are algorithmically cooled to a pure-state \cite{schulman1999molecular, boykin2002algorithmic}, which is then swapped with ancilla. We point out that since the noise channel (Eq.~\ref{eq:form_N}) always maps to a state which has a higher probability of being in $\ket{\alpha}$, it does not drive the auxillary qubits to the maximally mixed state and hence this cooling step is possible. The used auxillary qubits are then shifted to bring unused auxillary qubits next to the ancilla so that another RESTART gate can be performed when required.}

If the noise is assumed to act only before or after the unitary gates, then this shift operation can be performed without any errors. However, in the continuous-time setting, the noise can act \emph{while} the shift operation is being performed. Furthermore, since we could possibly need RESTART operations at $\Theta(\text{poly}(n))$ time, which would need $\Theta(\textnormal{poly}(n))$ shift operations --- thus, there is a possibility of accumulating a large error in the overall SHIFT operation at \emph{any}, no matter how small, non-zero $\kappa$. To resolve this issue, we propose to perform an imperfect shift operation, followed by allowing the noise to act on the shifted qubits for time $\tau_d$ to drive them to its fixed point [Fig.~\ref{fig:fig_2_4}(b)] . Clearly, if $\tau_d$ is chosen to be large enough, then the qubits would be in a state which can be subsequently cooled. However, increasing $\tau_d$ also increases the effective noise on the computational qubits since error correction is paused while the qubits are being restarted. A close analysis of this operation (supplement) reveals that to replenish $m$ auxillary qubits with the shift operation, $\tau_d$ can be chosen to be $\Theta(1 / \kappa^{1 - 1 / m})$ and hence the error sustained in the computational qubits while error correction is paused for this shifting, which is proportional to $\kappa \tau_d$, can be made smaller than the error correction threshold for sufficiently small $\kappa$. \hfill\(\square\)

 {Our next result considers purely dissipative dynamics --- if the noise channel is dephasing or amplitude damping, we provide theoretical evidence of the master equation remaining classically intractable at low noise rates. Our proof relies on using the Feynman clock construction~\cite{verstraete2009quantum}, to encode a fault-tolerant quantum computation in a local dissipative master equation albeit under postselection of clock qubits. Since postselected quantum circuits that can encode (postselected) quantum computations are unlikely to be classically tractable \cite{bremner2011classical,fujii2016noise, fujii2016computational}, we obtain the low-noise intractability of the noisy dissipative master equation.

\begin{theorem}
\label{theorem:dissipative_dyn}
{If $\mathcal{N}_i$ is the dephasing or amplitude damping channel, then for qubits arranged on two or higher dimensional lattices and for $\kappa$ below a threshold, there are instances of Eq.~\ref{eq:master_equation_paper} with $\mathcal{L}_0(t)$ being purely dissipative that cannot be weakly simulated on a classical computer within a small multiplicative error\footnote{A family of $n-$qubit quantum circuits is said to be weakly simulable within multiplicative error $c$ if a classical computer can be used to sample in $\textnormal{poly}(n)$ time within a probability distribution $p_\textnormal{cl}$ such that
\[
\frac{1}{c} p(x) \leq p_\textnormal{cl}(x) \leq c p(x) \ \forall \ x \in \{0, 1\}^n,
\]
where $p$ is the probability distribution at the output of the quantum circuit.} unless the polynomial hierarchy collapses to the third level.}
\end{theorem}
\emph{Proof sketch for theorem 3}. To dissipatively apply a unitary $U$ on $\rho_0$, we use an additional qubit, called the clock qubit, and $L = \ket{1}\bra{0}\otimes U$ --- with the initial state $\ket{0}\bra{0} \otimes \rho_0$ and postselecting on the clock qubit being in $\ket{1}$, the remaining qubits will be in $U \rho_0 U^\dagger$. Consider again $d = 2$ --- the computational qubits are laid out in one column, and the corresponding clock qubits are laid out in the rows [Fig.~\ref{fig:fig_2_4}(c)]. A fault-tolerant quantum circuit can now be encoded in the dissipative master equation with the unitaries encoded as shown above, and the RESTART operations encoded with just an amplitude damping channel. We show in the supplement that errors in both the computational qubits \emph{and} the clock qubits participating in a unitary can be translated to independent faults in the unitary gates being applied, and thus the threshold theorem still holds. {Finally, the clock qubits are replenished with a dissipative SHIFT to prepare for the next time-step in the circuit  (the SHIFT operation is performed again with two layers of SWAP, with SWAP being implemented dissipatively using the jump operators $\ket{0, 1}\bra{1, 0}, \ket{1, 0}\bra{0, 1}$) }--- we show in the supplement that if the noise channel under consideration is dephasing or amplitude damping, then the errors in the SHIFT operation do not impact the state of the computational qubits when postselected on the clock qubits being in $\ket{1}$. \hfill \(\square\)}

\edit{We remark that in theorem \ref{theorem:dissipative_dyn}, we assumed the ability to implement a purely dissipative Lindbladian for a chosen jump operator. Physically, due to lamb-shift and non-zero environment temperatures, a Lindbladian with jump operator $L(t)$ is accompanied with two corrections \cite{breuer2002theory, dann2018time} --- a Hamiltonian $\propto L^\dagger(t) L(t)$ (the lamb shift) and a Lindbladian with jump operator $L^\dagger(t)$ (the re-excitation). However, it can be shown that for the specific choice of the jump operators used above and with postselection on the clock qubits, these corrections do not impact the encoded quantum circuit. A detailed analysis of these corrections is provided in the supplement }

\emph{Conclusion}. We studied noisy dynamics of many-body open quantum spin systems with local interactions. Our work provides rigorous evidence of simulability transitions in their continuous-time dynamics. As specific technical problems, we leave open the extensions of theorems 2 and 3 to one-dimensional systems as well as to a larger class of noise channels. Furthermore, while we have exclusively focussed on Markovian spin systems, future directions could include studying simulability transitions in other experimentally relevant models of many-body quantum systems. These could include non-Gaussian bosonic systems, which would be a model for many quantum optics experiments, and non-Markovian quantum systems.

\begin{acknowledgements}
We thank Daniel Malz, Alavaro Alhambra and Georgios Styliaris for useful discussions and Cosimo Rusconi for providing feedback on the manuscript. RT acknowledges Max Planck Harvard research center for quantum optics (MPHQ) postdoctoral fellowship. We  acknowledge  support  from the  ERC  Advanced  Grant  QUENOCOBA  under  the EU  Horizon  2020  program  (grant  agreement  742102) and  from  the  Deutsche  Forschungsgemeinschaft  (DFG, German Research Foundation) under the project number414325145 in the framework of the Austrian Science Fund(FWF): SFB F7104.
\end{acknowledgements}

\onecolumngrid
\begin{center}
\large
\textbf{Supplementary material to \\ ``Simulability transitions in continuous-time dynamics of local open quantum systems" }
\end{center}
\tableofcontents
\section{Notation and prelimnaries}
{Given a finite-dimensional Hilbert space $\mathcal{H} \cong \mathbb{C}^d$, we will denote by $\mathfrak{D}_1(\mathcal{H})$ the set of density matrices over $\mathcal{H}$ i.e. $\rho \in \mathfrak{D}_1(\mathcal{H})$ if $\rho \geq 0$ (i.e. it is positive semi-definite) and $\text{Tr}[\rho] = 1$. We will denote by a $\mathfrak{L}(\mathcal{H})$ the vector space of linear operators mapping $\mathcal{H}$ to $\mathcal{H}$. Note that $\mathfrak{D}_1(\mathcal{H}) \subset \mathfrak{L}(\mathcal{H})$. A superoperator $\mathcal{M}$ on $\mathcal{H}$ is a linear operator mapping the space $\mathfrak{L}(\mathcal{H})$ to $\mathfrak{L}(\mathcal{H})$. Given a superoperator $\mathcal{M}$, its Choi-state $\Phi_\mathcal{M} \in \mathfrak{L}(\mathcal{H}\otimes \mathcal{H})$ is given by
\[
\Phi_\mathcal{M} = \big(\mathcal{M}\otimes \text{id}\big)(\ket{\Phi}\bra{\Phi}),
\]
where $\ket{\Phi}$ is the maximally entangled state over $\mathcal{H}\otimes \mathcal{H}$. A superoperator $\mathcal{L}$ is a Lindbladian if $\exists H, L_1, L_2 \dots L_m \in \mathfrak{L}(\mathcal{H})$, where $H = H^\dagger$ and $\text{Tr}[L_i] = 0$ for all $i \in \{1, 2 \dots m\}$, such that
\[
\mathcal{L}X = -i [H, X] + \sum_{j = 1}^m \bigg(L_j X L_j^\dagger - \frac{1}{2}\{L_j^\dagger L_j, X\}\bigg) \ \text{for all } X \in \mathfrak{L}(\mathcal{H}).
\]
$H$ is called the Hamiltonian for $\mathcal{L}$ and $L_1, L_2 \dots L_m$ will be called the jump operators for $\mathcal{L}$. We will always assume that the jump operators are traceless --- if not, then they can be made traceless by suitably redefining the Hamiltonian \cite{wolf2008dividing}. This representation of a Lindbladian is unique upto unitary linear combination of jump operators i.e. $H, L_1, L_2 \dots L_m$ and $H', L_1', L_2' \dots L_m'$ generate the same Lindbladian if and only if $H = H'$ and $L_i = \sum_{j = 1}^m U_{i, j} L_j'$ for some unitary $U \in \mathbb{C}^{m \times m}$.
A Lindbladian $L$ will be called \emph{purely Hamiltonian} if in this representation, $L_1, L_2 \dots L_m = 0$ and \emph{purely dissipative} if $H = 0$.

A superoperator $\mathcal{E}$ is a quantum channel, or a completely positive trace preserving map, if and only if $\exists E_1, E_2 \dots E_M \in \mathfrak{L}(\mathcal{H})$ satisfying $\sum_{i = 1}^M E_i^\dagger E_i = I$ such that
\[
\mathcal{E}(X) = \sum_{j = 1}^m E_i X E_i^\dagger \ \text{for all } X \in \mathfrak{L}(\mathcal{H}).
\]
Given a quantum channel $\mathcal{E}$, the superoperator $(\mathcal{E} - \textnormal{id})$ is a Lindbladian with Hamiltonian and jump operators given by
\[
H = i \sum_{j = 1}^m \bigg( \text{Tr}[E_j^\dagger] E_j - \text{Tr}[E_j] E_j^\dagger\bigg), \ L_j = E_j - \text{Tr}[E_j] \frac{I}{d}
\]
A quantum channel $\mathcal{E}$ is an entanglement breaking channel if it can be expressed as
\[
\mathcal{E}X = \sum_{i = 1}^l \sigma_i \text{Tr}[P_i X] \ \text{for all } X \in \mathfrak{L}(\mathcal{H}),
\]
where $\sigma_1, \sigma_2 \dots \sigma_l \in  \mathfrak{D}_1(\mathcal{H})$ are density matrices and $P_1, P_2 \dots P_l$ is a POVM (i.e.~$P_i \succeq 0$ and $\sum_{i = 1}^l P_i = I$). Of special interest are the entanglement breaking channels  depolarizing, dephasing and amplitude damping channels over qubits which are given by.
\begin{align*}
&\text{Depolarizing channel: }\mathcal{E}X = \text{Tr}[X] \frac{I}{2},\\
&\text{Amplitude damping channel: }\mathcal{E}X = \text{Tr}[X] \ket{0}\bra{0},\\
&\text{Dephasing channel: }\mathcal{E}X = \sum_{i \in \{0, 1\}} \text{Tr}[X \ket{i}\bra{i}] \ket{i}\bra{i}.
\end{align*}
Given two functions $f, g : [0, \infty) \to [0, \infty)$ of parameter $h$, the notation $f = O(g)$ as $h \to h_0 \implies \exists c > 0 : f(h) \leq cg(h)$ as $h \to h_0$. The notation $f = \Omega(g)$ as $h \to h_0 \implies \exists c > 0 : c g(h) \leq f(h)$ as $h \to h_0$. Finally, $f = \Theta(g)$ if $f = O(g)$ and $f = \Omega(g)$. Furthermore, $f = O(g_1) + O(g_2)$ as $h \to h_0$ if $\exists f_1, f_2$ such that $f = f_1 + f_2$ and $f_1 = O(g_1)$ as $h \to h_0$ and $f_2 = O(g_2)$ as $h \to h_0$.
}

\section{High noise regime}
In this section, we outline a proof of theorem 1 of the main text. We begin the section by recalling a basic result from the theory of subcritical percolation \cite{grimmett1997percolation, bazant2000largest} which provides a bound on the likelihood of obtaining a very large connected cluster in a (site) percolating lattice. We also provide a simple extension of this result to a percolation problem where the state of each site is dependent on the states of (some finite number of) its neighbouring sites --- this result is used for analyzing the run-time of the classical sampling algorithm used in the proof of theorem 1.

\subsection{Some results for subcritical percolation}
\begin{definition}[Cluster] Consider a percolation problem on a connected lattice $L \subset\mathbb{Z}^d$ where each vertex is either open (set to 1) or closed (set to 0). A set of closed vertices $S \subset L$ is called a cluster if $\forall v_1, v_2 \in S$, there exists a path (using only edges from $\mathbb{Z}^d$) from $v_1$ to $v_2$. We will denote by $\mathcal{C}(L)$ as the set of all clusters on $L$ i.e.~$\mathcal{C}(L) = \{S \subset L | S \text{ is a cluster}\}$.
\end{definition}
\begin{lemma} [Independent site-percolation from Ref.~\cite{bazant2000largest}]
\label{lemma:site_perc_problem}
Consider a site-percolation problem on a connected lattice $L \subset \mathbb{Z}^d$ with $n$ sites where each vertex $v$ is independently opened (set to 1) or closed (set to 0). Then, $\exists p_c \in (0, 1)$, referred to as the site percolation threshold, such that if $\textnormal{Prob}(X_v = 1) > p_c\ \forall \ v \in L$, where $X_v$ is the state of the vertex $v \in L$ then
\[
\textnormal{Prob}\bigg(\max_{S\in \mathcal{C}(L)} |S| \geq s\bigg) \leq O(n^d s^d \exp(-s/s_p)),
\]
for some $s_p > 0$.
\end{lemma}

Next, we consider a site percolation problem on $\mathbb{Z}^d$ where each site is statistically dependent on a vertices in a finite neighbourhood, and show that a similar percolation threshold can be derived for such a problem.

\begin{lemma}
\label{lemma:dependent_percolation}
Consider a site-percolation problem on a connected lattice $L\subset\mathbb{Z}^d$ with $n$ sites where each site is opened (set to 1) or closed (set to 0). The state of a vertex $v \in L$ being conditionally dependent on state of the vertices in a deleted neighbourhood\footnote{A deleted neighbourhood $N_v$ of a point $v$ on a lattice is a neighbourhood of the with the exclusion of $v$} $N_v \subset L$, then if $\inf_{v\in L}\inf_{x \in \{0, 1\}^{|N_v|}} \ \textnormal{Prob}(X_v = 1 | X_{N_v} = x) \geq p_c$, where for $v \in L, \ X_v \in \{0, 1\}$ is the state of the vertex $v$, for $S \in L$, $X_S \in \{0, 1\}^{|S|}$ is the state of the vertices contained in $S$ and $p_c$ is the site-percolation threshold defined in lemma \ref{lemma:site_perc_problem}, then 
\[
\textnormal{Prob}\bigg(\max_{S\in \mathcal{C}(L)} |S| \geq s\bigg) \leq O(n^d s^d \exp(-s / s_p)),
\]
for some $s_p >0$.
\end{lemma}
\noindent\emph{Proof:}  {The idea behind this proof is to construct a channel from a percolation problem where all sites are independent to the percolation problem where each site is dependent on its neighbours in such a way that the maximum cluster size can only decrease.} We consider a site-percolation problem on $L$ where each vertex $v \in L$ is independently open or closed. For notational convenience, $\forall  v \in L$, we define $x_v \in \{0, 1\}^{|N_v|}$ via
\[
x_v = \argmin_{x \in \{0, 1\}^{|N_v|}} \text{Prob}(X_v = 1 | X_{N_v} = x).
\]
Next, $\forall  v \in L$ and $x \in \{0, 1\}^{|N_v|}$, we define a Bernoulli random variable $Z_{v, x}$ such that
\begin{align}\label{eq:prob_z}
\text{Prob}(Z_{v, x} = 1) = 
\begin{cases} \text{Prob}(X_v = 1 | X_{N_v} = x_v) & \text{ if } x = x_v, \\
{ \big(\text{Prob}(X_v = 1 | X_{N_v} = x) - \text{Prob}(X_v = 1 | X_{N_v} = x_v)}\big)/ \text{Prob}(X_v = 0 | X_{N_v} = x_v) & \text{ if } x \neq x_v.
\end{cases}
\end{align}
The random variable $X_v$ can then be generated from $Z_{v, x}$ via
\[
X_v = \begin{cases}
1 & \text{ if } Z_{v, x_v} = 1, \text{ else } \\
Z_{v, x} & \text{ where }x \in \{0, 1\}^{|N_v|}\text{ such that } X_{N_v} = x.
\end{cases}
\]
 {In particular, we note that,
\begin{align*}
\text{Prob}\big(X_v = 1 | Z_{v, x'} = z_{x'} \ \text{for} \ x' \in\{0, 1\}^{\abs{N_v}}, X_{N_v} = x \big) = \begin{cases}
1 & \text{ if }z_{x_v} = 1 \text{ or } z_x = 1, \\
0 & \text{ otherwise}.
\end{cases}
\end{align*}
It is easy to verify that the definition of $X_v$ in terms of the independent random variables $Z_{v, x}$ for $x \in \{0, 1\}^{\abs{N_v}}$ produces the right conditional probability distributions. In particular, for $x \neq x_v$, we obtain that
\begin{align*}
&\text{Prob}(X_v = 1 | X_{N_v} = x) \nonumber\\
&\qquad = \sum_{z_{x'}, \ x' \in \{0, 1\}^{\abs{N_v}}} \text{Prob}\big(X_v = 1 | Z_{v, x'} = z_{x'} \ \text{for} \ x' \in\{0, 1\}^{\abs{N_v}}, X_{N_v} = x \big)  \prod_{x' \in \{0, 1\}^{\abs{N_v}}} \text{Prob}\big(Z_{v, x'} = z_{x'}\big) \nonumber\\
&\qquad = \text{Prob}(Z_{v, x_v} = 1) + \big(1 - \text{Prob}(Z_{v, x_v} = 1)\big) \text{Prob}(Z_{v, x} = 1),
\end{align*}
which is seen to be self-consistent using Eq.~\ref{eq:prob_z}. Similarly,
\begin{align*}
&\text{Prob}(X_v = 1 | X_{N_v} = x_v) \nonumber\\
&\qquad = \sum_{z_{x'}, \ x' \in \{0, 1\}^{\abs{N_v}}} \text{Prob}\big(X_v = 1 | Z_{v, x'} = z_{x'} \ \text{for} \ x' \in\{0, 1\}^{\abs{N_v}}, X_{N_v} = x_v \big)  \prod_{x' \in \{0, 1\}^{\abs{N_v}}} \text{Prob}\big(Z_{v, x'} = z_{x'}\big) \nonumber\\
&\qquad = \text{Prob}(Z_{v, x_v} = 1),
\end{align*}
which is again seen to be self-consistent using Eq.~\ref{eq:prob_z}.
}

Finally, we can easily note that  $Z_{v, x_v} = 1 \implies X_v = 1$. Therefore, the maximum cluster size in a lattice $L$ where each vertex $v$ has state $Z_{v, x_v}$ is necessarily larger than the maximum cluster size in a lattice $L$ where each vertex $v$ has state $X_v$. Thus, from lemma \ref{lemma:site_perc_problem}, the bound on the likelihood of the maximum cluster size follows. \(\hfill \square\)

\subsection{Sampling from the open-quantum spin system dynamics}
\noindent We provide a technically precise formulation of the continuous time problem considered in the main text.
\begin{problem}\label{prob:1}For a fixed lattice dimension $d \in \mathbb{N}$, an interaction range $R > 0$, an interaction strength $J$, a noise rate $\kappa > 0$ and a single spin entanglement-breaking channel $\mathcal{N}:\mathfrak{L}(\mathbb{C}^2) \to \mathfrak{L}(\mathbb{C}^2)$, sample in the computational basis from a family of states $\{ \rho_n \in \mathfrak{D}_1((\mathbb{C}^{2})^{\otimes n}): n \in \mathbb{N}\}$ where for $n \in \mathbb{N}$, $\rho_n := \rho(t_n)$ is a state of $n-$spins arranged on $\mathbb{Z}^d$ obtained from
\[
\frac{d\rho(t)}{dt} =\mathcal{L}_n(t) \rho(t) + \kappa \sum_{i = 1}^n \big(\mathcal{N}_i - \textnormal{id}\big)\rho(t),
\]
where
\begin{itemize}
\item $\forall n \in \mathbb{N}$, $\rho(0)$ is an $n$ spin product state that can be computed classically in $O(\textnormal{poly}(n))$ time,
\item $t_n = O(\textnormal{poly}(n))$ is the evolution time, 
\item $\forall n \in \mathbb{N}$ and $t \in [0, t_n]$, $\mathcal{L}_n(t):\mathfrak{L}((\mathbb{C}^{2})^{\otimes n}) \to \mathfrak{L}((\mathbb{C}^{2})^{\otimes n}) $ is an $n-$spin local Lindbladian superoperator which has an efficiently computable local representation of interaction range $R$ and interaction strength $J$.
\item $\forall i \in \{1, 2 \dots n\}$, $\mathcal{N}_i:\mathfrak{L}((\mathbb{C}^{2})^{\otimes n}) \to \mathfrak{L}((\mathbb{C}^{2})^{\otimes n}) $ is the tensor product of the entanglement breaking channel $\mathcal{N}$ acting on the $i^\text{th}$ spin and identity on other spins.
\end{itemize}\end{problem}

We first trotterize the evolution and represent each channel in the trotterized evolution (be it the entanglement breaking channel due to trotterization of the noise, or the channels that are entangling neighbouring qubits) as a convex combination of the identity channel and a completely-positive trace preserving (CPTP) channel. This is made precise in the following lemma.

 {\begin{lemma}
\label{lemma:trotter}
A parameter $N = \Theta(\textnormal{poly}(n) / \varepsilon)$ can be chosen such that $\norm{\rho_n - \hat{\sigma}_{n, N}}_1\leq \varepsilon$, where $\rho_n$ is defined in Problem \ref{prob:1} and $\hat{\sigma}_{n, N}$ is the trotterized state
\begin{align}\label{eq:state_to_sample_from}
\hat{\sigma}_{n, N} = \prod_{\tau = 1}^N \bigg[ \prod_{\Lambda } \bigg( \bigg(1 - \frac{gt_n}{N}\bigg)\textnormal{id} + \frac{gt_n}{N}\mathcal{E}^\Lambda_\tau \bigg)\bigg)\prod_{i = 1}^n\bigg(\bigg(1 - \frac{\kappa t_n}{N}\bigg)\textnormal{id} + \frac{\kappa t_n}{N}\mathcal{N}_i\bigg)\bigg)\bigg]\rho(0),
\end{align}
for some $g$ which depends on $(J, d, R)$, $\mathcal{E}_\tau^\Lambda$, which is a CPTP map acting on the qubits contained in $\Lambda$. The channel $(1 - gt / N) \textnormal{id} + gt / N \mathcal{E}_\tau^\Lambda$ appearing in the trotterized state, expressed as a convex combination of $\textnormal{id}$ and $\mathcal{E}_\tau^\Lambda$, will be referred to as a `horizontal' channel.
\end{lemma}
\noindent\emph{Proof:} We use the first order Trotter-Suzuki formula to approximate $\rho(t)$ by $\hat{\rho}_{t, N}$, where for $N \in \mathbb{N}$,
\begin{align}\label{eq:trotter_appx}
\hat{\rho}_{n, N} = \prod_{\tau = 1}^N \bigg[\prod_{\Lambda } \exp\bigg({\int_{(\tau - 1)t_n /N}^{\tau t_n / N} \mathcal{L}_n^\Lambda(s)ds}\bigg) \prod_{i = 1}^n \exp\bigg({\frac{\kappa t_n}{N} (\mathcal{N}_i - \textnormal{id})}\bigg)\bigg] \rho(0).
\end{align}
A standard analysis of the trotterization error \cite{childs2021theory} allows us to bound the trace-norm error between $\rho_n$ and $\hat{\rho}_{n, N}$:
\[
\norm{\rho_n - \hat{\rho}_{n, N}}_1 \leq O\bigg(\frac{t_n^2}{N}\bigg(\sum_{\Lambda} \sup_{s \geq 0}\norm{\mathcal{L}_n^\Lambda(s)}_{1\to 1} + n\kappa \norm{\textnormal{id} - \mathcal{N}}_{1\to 1}\bigg)^2\bigg).
\]
Since $\kappa = \Theta(1), \norm{\textnormal{id} - \mathcal{N}}_{1\to 1} = O(1)$ and $\forall \ \Lambda \in \mathcal{S}: \ \sup_{s\geq 0} \norm{\mathcal{L}_n^\Lambda(s)}_{1\to 1} \leq O(1)$, we obtain that $\norm{\rho(t) - \hat{\rho}_{n, N}}_{1} \leq O(t^2 \text{poly}(n) / N)$ --- a sufficiently large value of $N$ which scales as $\text{poly}(n)$ thus allows us to control the error incurred in this approximation.

Next, we consider each of the channels appearing in the trotterization (Eq.~\ref{eq:trotter_appx}) and express them as the convex combination of the identity channel and a completely-positive trace-preserving (CPTP) map. For a time-dependent Lindbladian $\mathcal{L}(s)$, consider the channel $e^{\int_0^\tau \mathcal{L}(s)ds}$ for some $\tau > 0$. A first order Taylor expansion of this map with respect to $\tau$ yields the map $\textnormal{id} + \int_0^\tau \mathcal{L}(s) ds$ --- note that if $\tau$ is sufficiently small (made precise below), this map is also CPTP. From Taylor's theorem it follows that
\[
\norm{\exp\bigg({\int_0^\tau\mathcal{L}(s)ds}\bigg) - \bigg(\textnormal{id} + \int_0^\tau \mathcal{L}(s) ds\bigg)}_{1\to 1} \leq O\bigg(\sup_{s\geq 0}\norm{\mathcal{L}(s)}_{1\to 1}^2 \tau^2\bigg).
\]
Next, we note that for any $g>0$, $\textnormal{id} + \int_0^\tau \mathcal{L}(s) ds = (1 - g\tau) \textnormal{id} + g\int_0^\tau(\textnormal{id} +  g^{-1}\mathcal{L}(s))ds$. Choosing $g =  \sup_{s\geq 0} \abs{\lambda_\text{max}(\Phi_{\mathcal{L}(s)})}$, where $\Phi_\mathcal{L}$ is the Choi-state corresponding to $\mathcal{L}$, it follows that $\int_0^\tau (\textnormal{id} + g^{-1}\mathcal{L}(s))ds$ is completely positive. Note that $g$ will be dependent only on $(J, d, R)$. Furthermore, since $\mathcal{L}$ is a Lindbladian, it follows that $\tau^{-1}\int_0^\tau\big(\textnormal{id} + g^{-1}\mathcal{L}(s)\big)ds$ is completely positive and trace preserving. For $\tau < 1 / g$, we have thus approximated $e^{\int_0^\tau\mathcal{L}(s)ds}$ by a convex combination of the identity channel (applied with probability $1 - g\tau$) and a non-identity CPTP channel (applied with probability $g\tau$). We therefore obtain that $\forall \Lambda$,
\[
 \exp\bigg({\int_{(\tau - 1)t_n /N}^{\tau t_n / N} \mathcal{L}_n^\Lambda(s)ds}\bigg)= \bigg(1 - \frac{g t}{N}\bigg) \textnormal{id} + \frac{gt}{N} \mathcal{E}^\Lambda_\tau + O\bigg(\frac{t^2}{N^2}\bigg)
\]
for some channel $\mathcal{E}_\tau^\Lambda$ that only acts on the qubits in $\Lambda$. Similarly, $\forall i \in \{1, 2 \dots n\}$:
\[
\exp\bigg({\frac{\kappa t_n}{N} (\mathcal{N}_i - \textnormal{id})}\bigg) = \bigg(1 - \frac{\kappa t_n}{N}\bigg) \textnormal{id} + \frac{\kappa t_n}{N} \mathcal{N}_i + O\bigg(\frac{t^2}{N^2}\bigg).
\]
We can then construct a circuit whose output is $\hat{\sigma}_{t, N}$, where
\begin{align}\label{eq:state_to_sample_from}
\hat{\sigma}_{n, N} = \prod_{\tau = 1}^N \bigg[ \prod_{\Lambda } \bigg( \bigg(1 - \frac{gt_n}{N}\bigg)\textnormal{id} + \frac{gt_n}{N}\mathcal{E}^\Lambda_\tau \bigg)\bigg)\prod_{i = 1}^n\bigg(\bigg(1 - \frac{\kappa t_n}{N}\bigg)\textnormal{id} + \frac{\kappa t_n}{N}\mathcal{N}_i\bigg)\bigg)\bigg]\rho(0).
\end{align}
Furthermore, it easily follows from the abovementioned error estimates that $\norm{\hat{\sigma}_{t, N} - \rho(t)}_1 \leq O(t^2 \text{poly}(n) / N)$ --- if $t = O(\text{poly}(n))$, then clearly there is a choice of $N = \Theta(\text{poly}(n) / \varepsilon)$ that ensures that $\norm{\hat{\sigma}_{t, N} - \rho(t)}_1 \leq \varepsilon$. \hfill\(\square\)}

Next, we consider the problem of sampling from $\hat{\sigma}_{n, N}$ --- we first map it to a percolation problem on a $(d + 1)$-dimensional lattice by choosing the identity channel, or the non-identity channel ($\mathcal{E}_\Lambda$ or $\mathcal{N}_i$). However, this sampling alone is not enough to ensure that the resulting problem percolates, since the probability of choosing the non-identity channel is very close to 0 --- in order to map this to a problem that percolates, each site in the percolation problem needs to be mapped to a block of several time-steps per qubit. This is made precise in the definition below and schematically depicted in Fig.~\ref{fig:supp_perc}.

\begin{definition} [Equivalent percolation problem]
\label{def:percolation}
Choose $\tau > 0$ and $m = \lceil{N \tau / t_n}\rceil$ --- a percolation problem on $\mathbb{Z}^{d+1}$ corresponding to sampling from $\hat{\sigma}_{n, N}$ in Eq.~\ref{eq:state_to_sample_from} is constructed by the first sampling from the convex combination of the channels ($\mathcal{E}_\Lambda$, $\mathcal{N}_i$ or the single- or two-qubit identity channels) in Eq.~\ref{eq:state_to_sample_from} --- then a vertex $v := (i, q) \in \mathbb{Z}^{d + 1}$ (where $i \in \mathbb{Z}^d$ and $q \in \mathbb{N}$) is open (i.e. $1$) if all the horizontal channels acting on the $i^\textnormal{th}$ qubit from time-step $mq$ to $m(q + 1) - 1$ are sampled to be the identity channel, and the channel $\mathcal{N}_i$ is applied at least once else it is closed (i.e.~$0$). 
\end{definition}

\begin{figure}[b]
\centering
\includegraphics[scale=0.75]{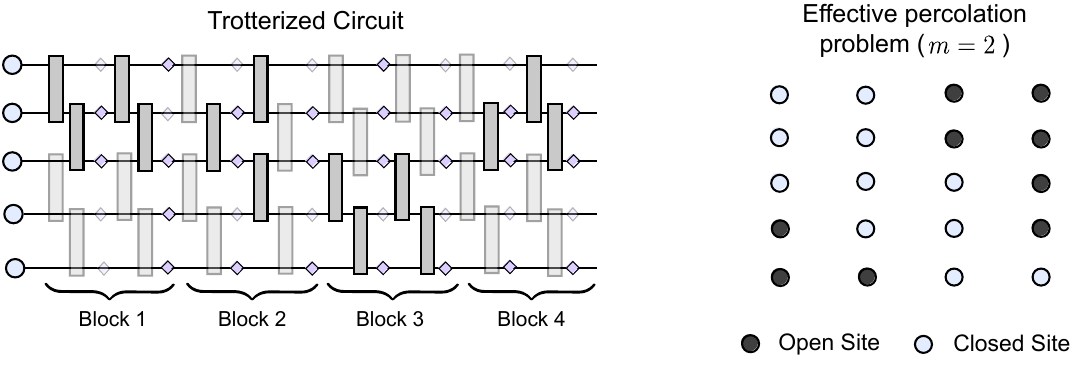}
\caption{An example of a sampled trotterized circuit in 1D for 5 qubits and 8 time-steps, which is mapped to a percolation problem as per definition \ref{def:percolation} with $m = 2$.}
\label{fig:supp_perc}
\end{figure}

\begin{reptheorem}{theorem:theorem_1} [Formal]
$\exists f_\textnormal{th} > 0$ dependent on $d$ and $R$ such that for $\kappa \geq f_\textnormal{th} J$, there is classical randomized polynomial-time algorithm to sample within an $\varepsilon$ total variation distance of $\rho_n$, the $n$ qubit state from the family of states specified in problem \ref{prob:1}, in time $O(\textnormal{poly}(n, 1 / \varepsilon))$.
\end{reptheorem}
\noindent \emph{Proof}: We next analyze the equivalent percolation problem and arrive at a proof of theorem 1 from the main text. Our goal is to show that for a sufficiently large $\kappa$, $\tau$ can be chosen to ensure that the equivalent percolation problem is subcritical which can then be efficiently sampled from. To do so we will use lemma \ref{lemma:dependent_percolation} --- we first provide a lower bound on the probability that a vertex on the lattice corresponding to the equivalent percolation problem is open, maximized over all possible configurations of the neighbouring vertices. We first fix some notation:
\begin{enumerate}
\item[(a)] For a vertex $v = (i, q)$ in the equivalent percolation problem, we associate the neighbourhood 
\[ N_v = \{( j \in \mathbb{Z}^d, s\in \mathbb{Z})\ |\ \exists \Lambda \text{ such that } i, j \in \Lambda\}.\] 
\item[(b)] For a set of vertices $\mathcal{V}$, we define by $\ell_\mathcal{V}$ the number of horizontal channels that involve the qubits associated with vertices in $\mathcal{V}$ qubit i.e.
\[
\ell_{\mathcal{V}} = \big| \{\Lambda\ |\ i \in \Lambda \text{ for some } (i, v) \in \mathcal{V}  \}\big |.
\]
\end{enumerate}
 {Next, given a vertex $v = (i, q)$, we consider lower bounding $\textnormal{Prob}(X_v = 1 | X_{N_v} = x)$, where $x \in \{0, 1\}^{\abs{N_v}}$ --- we do so by deriving a lower bound on $\textnormal{Prob}(X_v = 1,X_{N_v} = x)$, and an upper bound on $\textnormal{Prob}(X_{N_v} = x)$. Consider first $\textnormal{Prob}(X_v = 1,X_{N_v} = x)$ and note that the event $X_v = 1,X_{N_v} = x$ occurs if in the time-steps associated with $v$
 \begin{enumerate}
\item[(a)] All the horizontal channels involving the $i^\textnormal{th}$ qubit are sampled to be identity, 
\item[(b)] The noise channel is applied at least once on $i^\textnormal{th}$ qubit and all the qubits associated with vertices in $N_v$ which correspond to $1$ in $x$.
\item[(c)] The noise channel is not applied on all the qubits associated with vertices in $N_v$ which correspond to $0$ in $x$.
\end{enumerate}
Thus, we obtain the lower bound
\[
\textnormal{Prob}(X_v = 1, X_{N_v} = x) \geq \bigg(1 - \frac{gt}{N}\bigg)^{m \ell_{\{v\}}} \bigg(1 - \bigg(1 - \frac{\kappa t}{N}\bigg)^m \bigg)^{\norm{x}_1+1} \bigg(1- \frac{\kappa t}{N}\bigg)^{m(\abs{N_v} - \norm{x}_1)}
\]
Next, consider $\textnormal{Prob}(X_{N_v} = x)$ --- note that the event $X_{N_v} = x$ implies that one of the two events below has occured in the time-steps associated with $v$
\begin{enumerate}
\item[(a)] One horizontal channel associated with the qubits in $N_v$ is sampled to be identity.
\item[(b)] The noise channel is not applied on all qubits with vertices in $N_v$ which correspond to $0$ in $x$, and is applied at least once on all qubits with vertices in which correspond to $1$ in $x$. 
\end{enumerate}
From the union bound, we thus obtain
 \[
\text{Prob}(X_{N_v} = x) \leq \bigg(1 - \bigg(1 - \frac{\kappa t}{N}\bigg)^m \bigg)^{\norm{x}_1} \bigg(1- \frac{\kappa t}{N}\bigg)^{m(\abs{N_v} - \norm{x}_1)} + \bigg(1 - \bigg(1 - \frac{gt}{N}\bigg)^{m \ell_{N_v}}\bigg),
\]
Utilizing these estimates and substituting $m = \lceil N\tau / t\rceil$, we obtain that
\begin{align*}
\text{Prob}(X_v = 1|X_{N_v} = x) \geq e^{-g \tau \ell_{\{v\}}}(1 - e^{-\kappa \tau}) \bigg(\frac{e^{-\kappa \tau (\abs{N_v} - \norm{x}_1)}(1-e^{-\kappa \tau})^{\norm{x}_1}}{e^{-\kappa \tau (\abs{N_v} - \norm{x}_1)}(1-e^{-\kappa \tau})^{\norm{x}_1} + 1 - e^{-g \tau \ell_{N_v}}}\bigg) + O\bigg(\frac{1}{\text{poly}(N)}\bigg).
\end{align*}
We can now show that for $\kappa$ larger than a constant, the lower bound on $\inf_{x \in \{0, 1\}^{N_v}} \textnormal{Prob}(X_v = 1|X_{N_v} = x) $ can be made larger than the site-percolation threshold $p_c$. We note that due to the locality of interaction between the qubits, 
\[
\ell := \sup_v \ell_{\{v\}} \leq O(1)  \text{ and } \ell_N = \sup_v \ell_{N_v} \leq  O(1).
\]
Furthermore, $\abs{N_v} \leq R^d$ --- consequently, we obtain that
\[
\text{Prob}(X_v = 1|X_{N_v} = x) \geq  e^{-g\tau \ell}(1 - e^{-\kappa \tau}) \bigg(\frac{e^{-\kappa \tau (R^d - \norm{x}_1)}(1-e^{-\kappa \tau})^{\norm{x}_1}}{e^{-\kappa \tau (R^d - \norm{x}_1)}(1-e^{-\kappa \tau})^{\norm{x}_1} + 1 - e^{-g\tau \ell_{N}}}\bigg) + O\bigg(\frac{1}{\text{poly}(N)}\bigg).
\]
Now, fixing $\tau$ by $e^{-\kappa \tau} = c < 1$ for some constant $c$, we obtain that 
\[
\forall \ v: \inf_{x \in \{0, 1\}^{N_v}} \text{Prob}(X_v | X_{N_v} = x) \geq 1 - c + \min\big({c^{-R^d}}, {(1 - c)^{-R^d}}\big)O\bigg(\frac{g}{\kappa}\bigg) + O\bigg(\frac{1}{\text{poly}(N)}\bigg).
\]
Recall from lemma \ref{lemma:trotter} that $N = \Omega(\text{poly}(n))$ --- it thus follows that for large $n$, we can choose $\kappa / g$ to be large enough for the lower bound to be very close to $1 - c$ and thus an appropriate choice of $c$ (which is equivalent to an appropriate choice of $\kappa \tau$) will result in this probability being larger than the site-percolation threshold.} It then follows from lemma~\ref{lemma:dependent_percolation} that with a probability $1 - O(\text{polylog}(n) / \text{poly}(n))$, the maximum size of the percolating clusters will be $O(\text{log}(n))$. Using this fact, we can now provide an algorithm to sample approximately sample from the trotterized state $\hat{\sigma}_{t, n}$. We recall that an entanglement breaking single-qubit channel can be always be expressed as
\begin{align}\label{eq:entanglement_breaking_channel}
\mathcal{N}(\rho) = \sum_{i} \sigma_i \text{Tr}(E_i \rho),
\end{align}
where $\sigma_i \in \mathfrak{D}_1(\mathbb{C}^2)$, and $\{E_i\}$ form a POVM. Thus, an application of an entanglement breaking channel on a single-qubit state can be simulated by first measuring the qubit with respect to the POVM $E_i$ and then replacing the qubit with $\sigma_i$ if the outcome of the measurement is $i$. The sampling algorithm then proceeds in two steps:
\begin{enumerate}
\item[(a)] First, we sample from the percolation problem and ignore any samples that have a cluster of size larger than $\Omega(\text{log}(n)) + \Omega(\text{log}(1/\varepsilon))$. The error in total variation distance incurred due to this is $O(\varepsilon) $.
\item[(b)] We then replace the entanglement breaking channel by applying a POVM and replacing the qubit with an unentangled state as per Eq.~\ref{eq:entanglement_breaking_channel}. We note that this can be done efficiently and per each time-step. More specifically, suppose that uptil the (discrete) time-step $k$ all the entanglement breaking channels have been sampled in this way, and we wish to sample the entanglement breaking channels at time-step $k + 1$ --- we then need to sample from the measurement of the POVM corresponding to the qubits where the entanglement breaking channels are applied and trace over the remaining qubits. The sampling from the POVMs can be done sequentially --- we pick any one qubit and measure it in the POVM corresponding to the entanglement breaking channel, and trace over all the other qubits (including the ones on which the other entanglement breaking channel is applied). We note that we can efficiently contract the circuit to calculate the probabilities of various outcomes corresponding to the POVM since all the clusters of qubits can be contracted in $\text{poly}(n)$ time. After having computed these probabilities, we produce a sample from the POVM and conditioned on this sample we sample the next qubit which has an entanglement breaking channel. Since there are at-most $n$ such qubits, we need to do at most $\text{poly}(n)$ such contractions.
\end{enumerate}
The above two-steps outline a sampling algorithm that can be executed in $\text{poly}(n)$ time. Furthermore, there are two sources of error --- one is due to the trotterization, and the other is due to ignoring samples which have large clusters. Both of these errors are smaller than $O(\varepsilon)$ in the total-variation distance and this completes the proof of theorem 1. $\hfill\square$.



\section{Implication of threshold theorem on the low noise regime}

The threshold theorem in quantum computation \cite{aharonov2008fault} is a well known result which states that scalable quantum computation is possible if the noise rate in the quantum computer is below a threshold. These theorems are usually proved by performing encoded quantum computation together with error correction, and the reduction in the rate of noise needed to be able to implement scalable quantum computation depends on the error correcting code. While initially constructed and proved for general circuit model of computation, the threshold theorem has been proved for quantum circuits while constraining the interactions to be nearest neighbours on qubits arranged in 1D, 2D or 3D lattices \cite{aharonov2008fault, gottesman2000fault}.

Here, we investigate various implication of this threshold theorem on the low-noise classical intractability of open quantum dynamics. We begin by recalling a basic lemma proved by Aharanov and Gottesman, which indicates that a fault tolerant quantum computation is possible with 1D nearest neighbour qubits provided that qubits can be restarted at any point during the time evolution. This lemma is stated within the circuit model of computation with independent errors (although several extensions are possible) at different locations within the circuit with an error probability $\eta$.

\begin{lemma}[Fault tolerance in 1D with nearest neigbour gates from Ref.~\cite{aharonov2008fault}]
\label{lemma:aharanov_fault_tol}
Let $\varepsilon > 0$ and let $\mathcal{G}$ a universal set of gates. Let $Q$ be a quantum circuit on $n$ qubits and depth $\textnormal{poly}(n)$ formed from gates from $\mathcal{G}$ acting on either one or two neighbouring qubits. Then, $\exists \eta_\textnormal{th}$ such that for $\eta \leq \eta_\textnormal{th}$, there exists a circuit $C'$ formed from gates of $\mathcal{G}\cup\{\textnormal{SWAP}, \textnormal{RESTART}\}$ acting over $O(n\ \textnormal{polylog}(n / \varepsilon))$ qubits for depth $O(\textnormal{poly}(n) \textnormal{polylog}(n / \varepsilon))$ such that in the presence of noise $\eta \leq \eta_\textnormal{th}$, it computes a quantum state which is $\varepsilon$-close in total variation distance to the quantum state of the circuit $C$.
\end{lemma}

We point out that in the fault-tolerant construction used in this lemma, both the RESTART and SWAP operations can be subjected to errors, and the concatenated error correction schemes subsequently applied on them will succeed if these errors are small enough.

\subsection{Case: Lindbladian is purely Hamiltonian (non-dissipative)}
 {Building on this lemma and by providing implementations of the RESTART gates, Ref.~\cite{benquantum} showed that for the quantum circuits being interrupted with a constant rate of noise that is not depolarizing, a fault tolerant quantum computation can be implemented for sufficiently low noise. This had an implication of the low-noise regime of these circuits being classically hard to simulate. In the remainder of this section, we consider three different settings --- in the first two settings, which are in discrete-time, we review the construction of Ref.~\cite{benquantum} and then we extend it to the continuous time model.}
\begin{enumerate}
\item[(a)] For $d \geq 2$, we consider a discrete-time model where the entanglement breaking channels applied with a probability $p$ are interleaved with single and two-qubit unitaries applied locally on disjoint sets of qubits. Here, we show that as long as the entanglement breaking channel has a fixed point $\neq I / 2$, we expect the model to be classically intractable at sufficiently low rate of noise.
\item[(b)] For $d = 1$, we consider a discrete-time model where we apply a Matrix product unitary followed by the entanglement breaking channel applied with a probability $p$ on all qubits. Here, we again show that if the entanglement breaking channel has a fixed point $\neq I / 2$, we expect the model to be classically intractable at sufficiently low rate of noise.
\item[(c)] For $d \geq 2$, we consider a continuous-time model where the entanglement breaking channels can act at any time, and thus might result in erroneous gate operations. Here, we show that if the image of space of single qubit density matrices under the entanglement breaking channel does not contain $ I / 2$, then we expect this model to be classically intractable at sufficiently low rate of noise.
\end{enumerate}

A key ingredient that is used in the construction of Ref.~\cite{benquantum} is algorithmic cooling \cite{schulman1999molecular, boykin2002algorithmic} i.e.~having a supply of $m$ qubits individually being in a mixed state $\sigma_1 \otimes \sigma_2 \dots \otimes \sigma_m$ where $\sigma_i \neq I / 2 \ \forall \ i \in [m]$, we would like to be able to extract at least a single qubit in a pure state. This is a well known problem, and of technological relevance in NMR quantum computing where there is often a large supply of noisy qubits and it is of interest to concentrate the entropy on a few qubits. This will be a key ingredient in implementing the RESTART operations when the noise-channels do not necessarily drive the qubit state to identity (i.e.~they are not depolarizing). 

\begin{definition}[Polarization of a qubit state] A qubit (mixed) state $\sigma$ is said to have a polarization of $\varepsilon$ if $\lambda_\textnormal{max} - \lambda_\textnormal{min} = \varepsilon$, where $\lambda_\textnormal{max}$ and $\lambda_\textnormal{min}$ are the maximum and minimum eigenvalues of $\sigma$ respectively.
\end{definition}

\begin{lemma}[Algorithmic cooling from Ref.~\cite{schulman1999molecular}]
\label{lemma:alg_cooling}
Consider $m$ qubits initially in the state $\rho_\textnormal{init} = \bigotimes_{i=1}^m \sigma_i$, where for $i \in [m]$, $\sigma_i$ has polarization $\varepsilon_i (>0)$ and let $\varepsilon_\textnormal{min} = \min_{i \in [m]} \varepsilon_i > 0$, then for any $\eta \in (0, 1)$, $m$ can be chosen as a function of $\eta, \varepsilon_\textnormal{min}$ such that $\exists$ a unitary circuit $U$ of depth dependent on $\eta, \varepsilon_\textnormal{min}$ which yields (at least) one qubit whose reduced state has polarization $1 - \eta$.
\end{lemma}

\begin{figure}
\centering
\includegraphics[scale=0.65]{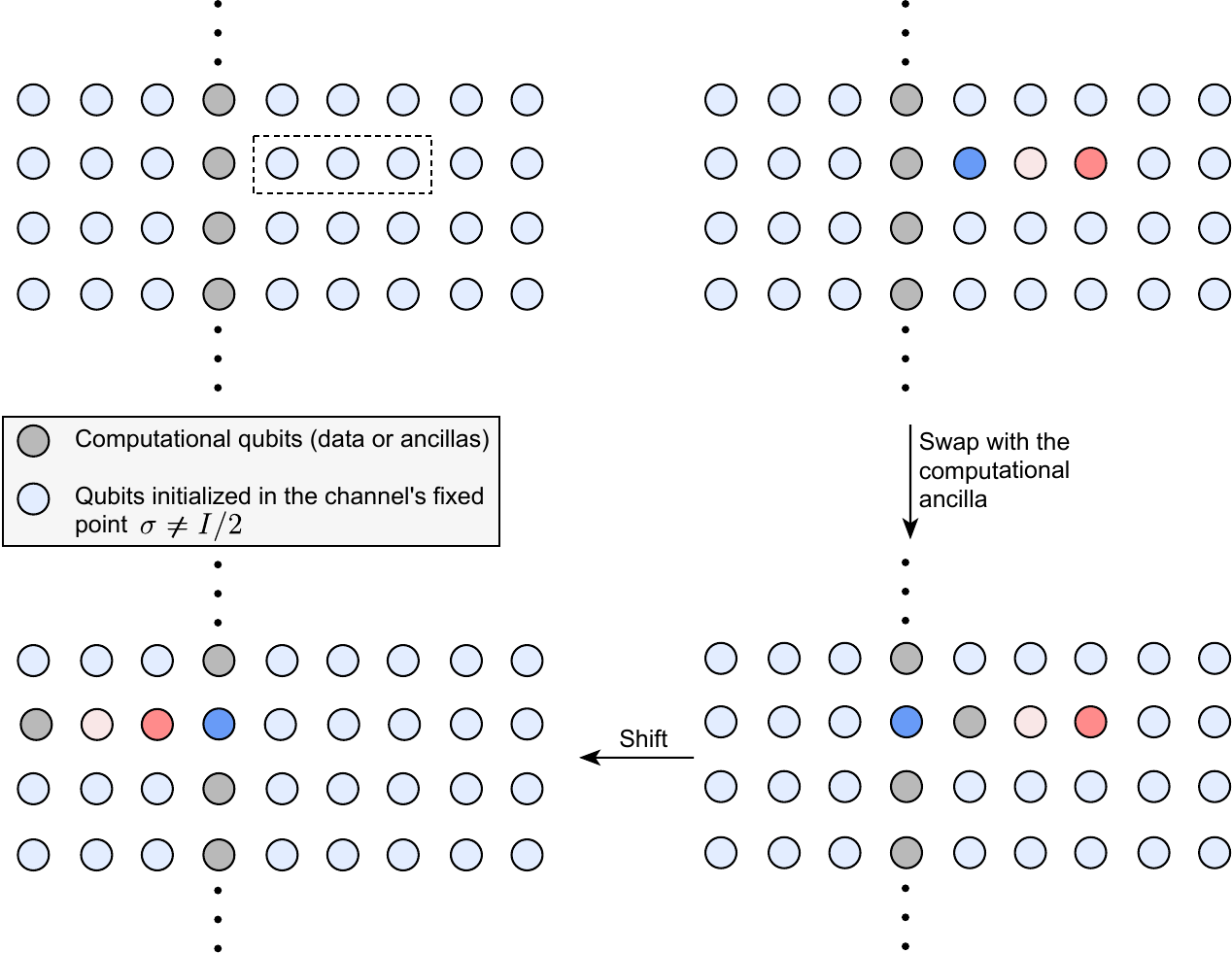}
\caption{A schematic depiction of the implementation of the RESTART operation on a 2D lattice of qubits. We note that the a shift operation can be implemented on the alternate qubits in any row (i.e.~the qubits at the odd or even positions) by doing two consecutive swap operations --- one layer of operations swapping the even with the odd qubits and then another layer of operations swapping the odd qubits with the even qubits. Taken together, these two operations will shift the odd qubits to one direction and the even qubits in the opposite direction. For simplicity, in this figure, we depict this shift operation as a single operation and hide one set of alternate qubits.}
\label{fig:restart_schematic}
\end{figure}

 {In the next two propositions, we consider first the discrete-time model of quantum computation in order to illustrate the basic construction behind the proof of low-noise BQP-completeness of the continuous-time model. These two propositions follow directly from the analysis presented in Ref.~\cite{benquantum}, but we include a proof of this for completeness. We consider two types of unitary layers in the discrete-time model --- a parallelized layer of unitary gates acting on disjoint sets of qubits, and matrix product unitary layers \cite{cirac2017matrix}.
\begin{definition}
\label{def:loc_par_unit}
A superoperator $\mathcal{U}:\mathfrak{L}((\mathbb{C}^{2})^{\otimes n})\to \mathfrak{L}((\mathbb{C}^{2})^{\otimes n})$ over $n$ spins arranged on $\mathbb{Z}^d$ is a local parallelized unitary layer if  $\forall \rho \in \mathfrak{L}((\mathbb{C}^{2})^{\otimes n})$
\[
\mathcal{U}\rho = \bigg(\prod_{i=1}^M U_i \bigg) \rho \bigg(\prod_{i=1}^M U_i^\dagger \bigg),
\]
for some $U_1, U_2 \dots U_M$ that are single or nearest-neighbour two-qubit unitaries acting on disjoint sets of spins.
\end{definition}

\begin{definition}
\label{def:mpu_unit}
A superoperator $\mathcal{U}:\mathfrak{L}((\mathbb{C}^{2})^{\otimes n})\to \mathfrak{L}((\mathbb{C}^{2})^{\otimes n})$ over $n$ spins is a matrix-product unitary layer of bond dimension $D$ if $\forall \rho \in \mathfrak{L}((\mathbb{C}^{2})^{\otimes n})$, $\mathcal{U}\rho = U \rho U^\dagger$, where $U$ is a $n$ qubit unitary such that for $\forall i_1, i_2 \dots i_n, j_1, j_2 \dots j_n \in \{0, 1\}$
\[
\bra{i_1, i_2 \dots i_n}U\ket{j_1, j_2 \dots j_n} = \prod_{l=1}^n A_l^{i_l, j_l},
\]
where $\forall i_1, i_2 \dots i_n, j_1, j_2 \dots j_n \in \{0, 1\}$, $A_1^{i_1, j_1} \in \mathbb{C}^{1 \times D}, A_n^{i_n, j_n} \in \mathbb{C}^{D \times 1}$ and $A_2^{i_2, j_2}, A_3^{i_3, j_3} \dots A_{n - 1}^{i_{n-1}, j_{n - 1}} \in \mathbb{C}^{D\times D}$.
\end{definition}
We point out that every local parallelized unitary layer in $d = 1$ dimensions can be expressed as a matrix product unitary layer with bond-dimension 2, but there are matrix product unitary layers that cannot be expressed as a local parallelized unitary layers.

\begin{problem}[Discrete time model]
For a fixed noise rate $p \in (0, 1)$ and a single spin entanglement-breaking channel $\mathcal{N}:\mathfrak{L}(\mathbb{C}^2) \to \mathfrak{L}(\mathbb{C}^2)$, sample in the computational basis from a family of states $\{\rho_n \in \mathfrak{D}_1((\mathbb{C}^2)^{\otimes n}): n \in \mathbb{N}\}$ where for $n \in \mathbb{N}$, $\rho_n$ is a state of $n$ spins given by
\[
\rho_n = \mathcal{E}_p^{\otimes n} \mathcal{U}_{t_n} \mathcal{E}_p^{\otimes n}  \mathcal{U}_{t_n - 1} \dots \mathcal{U}_1 \rho(0),
\]
where
\begin{itemize}
\item $\forall n \in \mathbb{N}$, $\rho_n(0)$ is a $n$ spin product state that can be computed classically in $O(\textnormal{poly}(n))$ time,
\item $t_n = O(\textnormal{poly}(n))$ is the number of time steps,
\item $\forall n \in \mathbb{N}, t \in \{1, 2 \dots t_n\}$, $\mathcal{U}_t$ is a unitary layer,
\item $\mathcal{E}_p = (1 - p) \textnormal{id} + p\mathcal{N}$.
\end{itemize}
\end{problem}

\begin{proposition}\label{theorem:discrete_time_circuit}
If $\exists \sigma \neq I / 2$ such that $\mathcal{N}(\sigma) = \sigma$, then $\exists \ p_\textnormal{th} > 0$ such that for $p \leq p_\textnormal{th}$ problem 2 with local parallelized unitary layers in two or higher dimensions (definition \ref{def:loc_par_unit}) cannot be weakly simulated on a classical computer within a total variation error\footnote{A family of $n-$qubit quantum circuits is said to be weakly simulable within $\varepsilon$-total variation error if a classical computer can be used to sample in $\textnormal{poly}(n)$ time within a probability distribution $p_\textnormal{cl}$ such that $\norm{p - p_\textnormal{cl}}_1 \leq \varepsilon$, where $p$ is the probability distribution at the output of the quantum circuit.} $\in (0, 1/2)$ unless $\textnormal{BQP} = \textnormal{BPP}$.
\end{proposition}
\noindent\emph{Proof}: We restrict ourselves to $d = 2$ (i.e.~a 2D lattice of spins). We show that any local 1D quantum circuit can be fault-tolerantly encoded into the dynamics of a 2D circuit with local parallelized unitary layers --- since there exist 1D quantum circuits that can solve BQP-complete problems, this implies the weak classical intractability of problem 2 with local parallelized unitary layers within a total variation error $\in (0, 1/2)$. We will implement the 1D fault-tolerant quantum computation (lemma ~\ref{lemma:aharanov_fault_tol}) in one of the columns of the 2D lattice, and call the qubits involved in this construction as the \emph{computational qubits} --- some of the computational qubits will be the data qubits (which encode the quantum computation) and some of these qubits will be ancilla qubits needed for the error correction operations. In order for the fault-tolerant construction to succeed, we need to implement a RESTART gate, with fidelity below the fault tolerance threshold, on the ancilla qubits. In order to restart a computational ancilla qubit, we use the non-computational qubits in its row. A schematic depiction of the restart operation is shown in Fig.~\ref{fig:restart_schematic} --- we start off with all the qubits to the right of the qubit to be restarted in a fixed point $\sigma \neq I / 2$ of the noise channel. We then perform a sequence of three steps:
\begin{enumerate}
\item First, we perform algorithmic cooling on certain number of, say $m$, qubits on the right of the qubit to be restarted. At the end of this step, the qubits neighbouring the computational ancilla qubit will be in a state $\ket{0}$. We note that the algorithmic cooling operation can fail with some non-zero probability --- however, this probability does not scale with the number of qubits involved and hence this construction still yields a threshold theorem. We analyze this point in more detail below.
\item Next, we swap the cooled qubit with the computational qubit. This effectively implements the RESTART operation.
\item Finally, we prepare the qubits for another RESTART operation - we note that the the previous two steps left the qubits immediately on the right of the computational qubit in a state that is no longer $\sigma$ and consequently cannot be used to implement another RESTART operation. We therefore now perform a shift operation to shift these qubits to the left of the computational qubits and replace the $m$ qubits on the right of the computational qubits with the state $\sigma^{\otimes m}$. This configuration can now be used to implement the next RESTART operation by using steps 1-3.
\end{enumerate}
\emph{Error analysis}: We need to ensure that a threshold theorem exists with the above outlined RESTART operation. Recall that at every time-step, we are allowed to apply either a single-qubit or a nearest neighbour two-qubit gate over disjoint set of qubits. Recall from lemma~\ref{lemma:alg_cooling} that to achieve a cooled qubit in the state $\ket{0}$ with specified probability $p_\text{cool}$, the number of qubits $m$ required is a function of $p_\text{cool}$ (which we denote by $m(p_\text{cool})$), and the number of time-steps $t$ required to accomplish this is also a function of $p_\text{cool}$ (which we denote by $t_\text{cool}(p_\text{cool})$). Furthermore, it follows from lemma~\ref{lemma:alg_cooling} that to achieve $p_\text{cooling} = 1 - O(1)$, both $m(p_\text{cool}) = O(1)$ and $t_\text{cool}(p_\text{cool}) = O(1)$. Consequently, both the following swap and shift operations (which needs to swap $O(m(p_\text{cool}))$ qubits) can be done in time-steps $t_\text{swaps}(p_\text{cool}) = O(1)$. Importantly, we note that in the model being considered in this proposition, the error channels are applied before or after the unitary layers. Consequently, we can perform the SWAP gate without an error \emph{in the gate}, and since $\sigma$ is a fixed-point of the noise channel, swapping a qubit in the state $\sigma$ with another qubit will still generate an output qubit in the state $\sigma$ (this is not necessarily true if an error occurs in the gate) and consequently the shift operation can always be performed to arrange the qubits properly for the next RESTART operation. Finally, since the number of time-steps and the number of qubits participating in the RESTART operation $O(1)$, we note that the RESTART operation can be made more efficient than required by the error correction threshold by making the probability of applying the entanglement breaking noise $p$ small enough. This shows that below a particular noise threshold, we will be able to implement a BQP hard problem fault-tolerantly within the discrete-time model considered, which proves the theorem. \hfill\(\square\).\\

\begin{proposition}
\label{theorem:1d_mpu}
If $\exists \sigma \neq I / 2$ such that $\mathcal{N}(\sigma) = \sigma$, then $\exists \ p_\textnormal{th} > 0$ such that for $p \leq p_\textnormal{th}$, problem 2 with matrix-product unitary layers of bond dimension $\geq 2$ (definition \ref{def:mpu_unit}) cannot be weakly simulated on a classical computer within a total variation error $\in (0, 1/2)$ unless $\textnormal{BQP} = \textnormal{BPP}$.
\end{proposition}
\noindent \emph{Proof}: This key difficulty with implementing a RESTART operation in the 1D nearest-neighbour setting is that the qubits that are cooled and swapped into the computational ancillas cannot be placed close to the computational ancillas, and $\Theta(n)$ (where $n$ is the number of computational qubits) swap operations are required in order to bring them close to the computational ancillas. Applying $\Theta(n)$ swap gates without any error correction breaks the threshold theorem for quantum computation and hence does not allow us to fault tolerantly encode a BQP hard problem into the noisy quantum dynamics. However, if we allow ourselves to use constant bond dimension MPUs, then we can perform $\Theta(n)$ swap operations with an MPU of bond dimension 2 in one time-step, and thus the construction used in the proof of theorem 2 works. In this case, we will just maintain the qubits used for restarting the computational ancilla on one end of the 1D lattice, and cool and swap them with the computational ancilla qubits in a single time-step.  \hfill\(\square\) \ \\}

\noindent We now proceed to the proof of theorem 2 where we consider a continuous model of noise being applied on the spins. Here, the swap operations needed to shift the qubits to be cooled near the computational qubits cannot be done faultlessly. However, if we consider noise modelled with point channels with a fixed point different from identity, we show that a 1D fault tolerant quantum computation can still be implemented in $d\geq 2$. \ \\

\begin{figure}[b]
\centering
\includegraphics[scale=0.7]{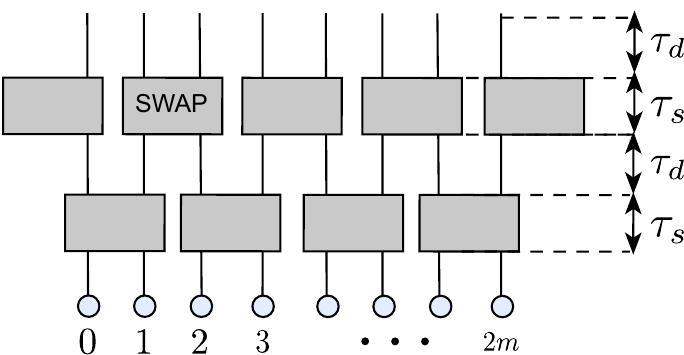}
\caption{Schematic of the shift operation used in the proof of theorem 4.}
\label{fig:shift}
\end{figure}
\begin{reptheorem}{theorem:cont_time_circuit}[Formal]
If $\mathcal{N}$ is of the form described in Eq.~4 of the main text, then for qubits arranged on two or higher dimensional lattices, $\exists \kappa_\textnormal{th}>0$ such that for $\kappa \leq \kappa_\textnormal{th}$ problem \ref{prob:1} with purely non-dissipative Lindbladians cannot be weakly simulated on a classical computer within total variation error $\in (0, 1/2)$ unless $\textnormal{BQP\ =\ BPP}$.
\end{reptheorem}

\noindent \emph{Proof}: For the two (or higher) dimensional models where the noise is applied continuously in time, one of the key difficulties in the construction of theorem 2 is that the SWAP operations can be faulty. Thus, even at a very low rate of noise, it is not guaranteed that the qubits that are being used to restart the computational ancillas at times $\Theta(\text{poly}(n))$ are in a state that is sufficiently different from the identity. In particular, if we consider channels which have identity as a fixed point (in addition to fixed points which are not identity) then an error in the SWAP gate could result in one of the qubits being set into the identity and thus not be useful for the RESTART operation.

However, channels satisfying Eq.~4 have the property that it maps every qubit state to a state of the form 
\[
\frac{(1 + \varepsilon)}{2} \ket{\alpha}\bra{\alpha} + \frac{(1 - \varepsilon)}{2}\ket{\beta}\bra{\beta},
\]
where $\varepsilon \geq 2 \lambda_\text{min}(P) - 1 > 0$ independent of the input state. 
Leveraging this property of $\mathcal{N}$, we can construct a shift operation that, despite being erroneous, ensures with a sufficiently high success probability that the shifted qubits are unentangled and different from identity and hence can be used for the subsequent cooling algorithm. This shift operation, implemented with two layers of SWAP gates, is shown in Fig.~\ref{fig:shift}  --- we apply the Hamiltonian for the SWAP gate for a time period $\tau_s$ (chosen such that in a noiseless setting, the SWAP gate would be perfectly executed) and then allow the qubits to evolve just under the influence of the entanglement breaking noise channel for a time period $\tau_d$. The latter evolution not only disentangles the output qubit due to the entanglement breaking nature of the channel, but also drives it away from $I / 2$. We make this concrete by analyzing the operation of this shift operation on $2m + 1$ qubits, out of which with probability $1 - q$, the qubits $2, 4, 6 \dots 2m$ are in a separable state with the states of the individual qubits belonging to $\mathcal{N}(\mathfrak{D}_1(\mathbb{C}^2))$ i.e.
\[
\rho(0) = (1 - q) \rho_{0, 1, 3, 5 \dots 2m - 1} \otimes \sigma_{2, 4, 6 \dots 2m}+ q\rho'.
\]
where $\sigma_{2, 4, 6 \dots 2m}$ can be expressed as
\begin{align}\label{eq:separable_state}
\sigma_{2, 4, 6 \dots 2m} = \sum_{\alpha} p_\alpha \sigma_2^\alpha \otimes \sigma_4^\alpha \dots \otimes \sigma_{2m}^\alpha,
\end{align}
with $\sum_{\alpha} p_\alpha = 1, p_\alpha \geq 0$ and $\sigma_i^\alpha \in \mathcal{N}(\mathfrak{D}_1(\mathbb{C}^2)) \ \forall \alpha, i \in \{2, 4, 6 \dots 2m\}$.
Note that with probability $\exp(-2m\kappa \tau_s)$, the SWAP operation occurs perfectly and hence the output state can be expressed as.
\[
\rho(\tau_s) = \exp(-2m\kappa \tau_s)(1 - q) \sigma_{1, 3, 5 \dots 2m - 1} \otimes \rho'_{0, 2, 4 \dots 2m} + (1 - \exp(-2\kappa \tau_s)(1 - q)) \rho'',
\]
where $\sigma_{1, 3, 5 \dots 2m - 1}$ is the state $\sigma_{2, 4, 6 \dots 2m}$ but now shifted to the qubits $1, 3, 5 \dots 2m - 1$ and $\rho'_{0, 2, 4 \dots 2m}$ ($\rho''$) is some state over the qubits $0, 2, 4 \dots 2m$ (all the qubits). Next, we allow the qubits to evolve under the entanglement breaking noise for a time $\tau_d$ --- we note that this is equivalent to applying the channel $\mathcal{E}_{\tau_d} = e^{\kappa \tau_d(\mathcal{N} - \text{Id})}$ on each qubit. By a Taylor expansion of this exponential, it can easily be seen that
\[
\mathcal{E}_{\tau_d} = e^{-\kappa \tau_d} \text{Id} + (1 - e^{-\kappa \tau_d}) \mathcal{N}',
\]
where $\mathcal{N}':\mathfrak{L}(\mathbb{C}^2) \to \mathfrak{L}(\mathbb{C}^2)$ is another entanglement breaking channel with the property that $\mathcal{N}'(\mathfrak{D}_1(\mathbb{C}^2)) = \mathcal{N}(\mathfrak{D}_1(\mathbb{C}^2))$. Therefore, with probability $1 - e^{-\kappa \tau_d}$, this channel disentangled the qubit and maps it to a state in $\mathcal{N}(\mathfrak{D}_1(\mathbb{C}^2))$ and consequently,
\[
\rho(\tau_s + \tau_d) = \mathcal{E}_{\tau_d}^{\otimes (2m + 1)}\rho(\tau_s) = (1 - q') \sigma_{0, 2, 4 \dots 2m - 2} \otimes \rho''_{1,3,5 \dots 2m-1, 2m} + q' \rho''',
\]
where $1 - q' = \exp(-2m\kappa \tau_s) (1 - q) + (1 - \exp(-\kappa \tau_d))^m(1 - \exp(-2\kappa \tau_s)(1 - q))$, $\sigma_{0, 2, 4 \dots 2m - 2}$ is a separable state of the form of Eq.~\ref{eq:separable_state} but over the qubits $0, 2, 4 \dots 2m - 2$. It immediately follows that $q' \leq q$ if
\begin{align}\label{eq:choice_tau_d}
\tau_d \geq -\frac{1}{\kappa}\log\bigg(1 - \bigg(\frac{(1 - q)(1 - e^{-2m \kappa \tau_s})}{1 - (1 - q)e^{-2m\kappa \tau_s}} \bigg)^{1/m}\bigg)
\end{align}
A similar condition holds for the next layer of SWAP gates --- if $\tau_d$ is chosen as per Eq.~\ref{eq:choice_tau_d}, then the final state of the $m$ qubits being shifted will be maintained in a seperable state where the individual qubits are in $\mathcal{N}(\mathfrak{D}_1(\mathbb{C}^2))$ with a probability at least $1 - q$ --- hence this property will also be maintained by the subsequent shift operation and we will have $m$ qubits available for the cooling procedure. We note that even though the qubits now being purified are in a separable state as opposed to a product state, since by assumption the eigenstates of the individual qubit states are the same, the cooling unitary applied on the product state will also cool the separable state without any impact on the fidelity of the cooling algorithm. However, the cooling algorithm's fidelity will be lowered by a factor $\geq 1 - q$. Consequently, in order to implement the RESTART operation with fidelity large enough to ensure that the threshold theorem can be satisfied, we must be able to choose a sufficiently small $q$. Since the shift operation outlined above relies on the entanglement breaking channel to purify the state of the qubits, it is not clear if this condition can be satisfied (for small rates $\kappa$) while at the same time ensuring that the noise in the computational qubits is smaller than the fault-tolerance threshold. However, this is indeed possible --- to see this, we note that for small $\kappa$, we can use $\tau_d$
\[
\tau_d \approx \bigg(\frac{2m(1 - q)}{q}\bigg)^{1/m} \frac{(\kappa \tau_s)^{1/m}}{\kappa}.
\]
However, note that for the error correction procedures to fault tolerantly encode a computation on the computational qubit, it is necessary to ensure that $\kappa \tau_d \leq c_\text{th}$, where $c_\text{th}$ is a constant dependent on the spread of the error correcting code. It follows easily that for a given $q$ and $m$ (which are determined by the target fidelity of the cooling operation) if $\kappa < qc_\text{th}^m / 2m\tau_s ( 1 -q)$, then for any $\tau_d \in (({2m(1 - q)}/{q})^{1/m} \kappa^{1/m - 1} \tau_s^{1/m}, c_\text{th} / \kappa)$, the shift operation will successfully bring the auxillary qubits near the computational qubits for the RESTART operation while at the same time ensuring that a quantum computation can be fault tolerantly performed on the computational qubits. \hfill\(\square\)

\subsection{Case: Purely dissipative lindbladian}
\subsubsection{Construction of a classically intractable instance}
\noindent  {Here, we consider the purely dissipative master equation in the presence of entanglement breaking noise,
\begin{align}\label{eq:dissipative_master_eq_w_noise}
\frac{d\rho(t)}{dt} =  \sum_{i = 1}^M\bigg( L_i(t)\rho L_i^\dagger(t) - \frac{1}{2}\{\rho, L_i^\dagger(t) L_i(t)\}\bigg) + \kappa \sum_{i = 1}^N \big(\mathcal{N}_i \rho - \rho \big).
\end{align}
where $\forall t, \textnormal{tr}[L_i(t)] = 0$. It is known that in the absence of noise (i.e.~for $\kappa = 0$), the dissipative lindbladian can be chosen to encode a quantum circuit on $n$ qubits with depth $ \textnormal{poly}(n)$ into its fixed point and the fixed point is reached in time $\textnormal{poly}(n)$. Hence, it is unlikely that Eq.~\ref{eq:dissipative_master_eq_w_noise} can be simulated efficiently on a classical computer when $\kappa = 0$. Below, we show that when the single-qubit noise channel $\mathcal{N}_i$ is either amplitude-damping or dephasing, this property is retained for a small but constant $\kappa$ if we restrict ourselves to local dissipative Linbladians in 2 or higher dimensions.

We show how to encode a quantum computation in a local master equation of this form but with post selection. We first consider the problem of applying a single unitary on a set of qubits and use the Feynman clock trick, similar to that used in Ref.~\cite{verstraete2009quantum}, to apply this unitary using an ancillary clock qubit. In our analysis, we additionally consider the impact of errors on both the clock and data qubits and show that for the dephasing and amplitude damping channel, this translates to effective local faults of the form considered in Ref. in the quantum computation.
\begin{lemma}[Faults in dissipatively encoded unitaries]
\label{lemma:deph_unit}
Consider a single-qubit noise channel $\mathcal{N}$ which satisfies
\[
\mathcal{N} \rho = \sigma_0 \bra{0}\rho\ket{0} + \sigma_1 \bra{1}\rho \ket{1} \ \forall \rho \in \mathfrak{D}(\mathbb{C}^2),
\]
where $\sigma_0 = (1 - p_0) \ket{0}\bra{0} + p_0 \ket{1}\bra{1}$ and $\sigma_1 = p_1 \ket{0} \bra{0} + (1 - p_1) \ket{1}\bra{1}$ for some $p_0, p_1 \in (0, 1)$.
Given $k$ data qubits in an initial state $\rho_0$ and a unitary $U$, then the dissipative master equation on $k + 1$ qubits (one clock qubit and $k$ data qubits) with jump operator $L =\ket{1}\bra{0} \otimes U $ on evolving an initial state $ \ket{0}\bra{0} \otimes \rho_0$ for $t = \Theta(1)$ and postselecting the clock qubit to be in $\ket{1}$ implements a $k-$qubit channel $\mathcal{E}$ such that
\[
\norm{\mathcal{E} - \mathcal{U}}_{1\to 1} \leq O\big({k\kappa} \big),
\]
where $\mathcal{U}\rho = U \rho U^\dagger$.
\end{lemma}
\noindent\emph{Proof}: We observe that since $\textnormal{span}(\ket{0}\bra{0}, \ket{1}\bra{1})$ remains invariant under the action of $\mathcal{N}$, the master equation (Eq.~\ref{eq:dissipative_master_eq_w_noise}) with the noise channel $\mathcal{N}$, if the initial state of the $k + 1$ qubits is $\ket{0}\bra{0}\otimes \rho_0$ then the state at time $t$, $R(t)$ can be expressed as
\[
R(t) = \ket{0}\bra{0} \otimes R_0(t) + \ket{1}\bra{1}\otimes R_1(t),
\]
where $R_0(t), R_1(t)$ are positive semi-definite $k-$qubit operators. It follows from the master equation that,
\begin{subequations}\label{eq:consequence_of_master_eq}
\begin{align}
&\frac{d}{dt}R_0(t) = -p_0\kappa R_0(t) + p_1 \kappa R_1(t) - R_0(t) + \kappa \sum_{i = 1}^k \big(\mathcal{N}_i -\textnormal{id}\big) R_0(t),\\
&\frac{d}{dt} R_1(t) =  -p_1\kappa R_1(t) + p_0\kappa R_0(t) + U R_0(t) U^\dagger + \kappa \sum_{i = 1}^k \big(\mathcal{N}_i -\textnormal{id}\big) R_1(t).
\end{align}
\end{subequations}
We note that 
\begin{subequations}
\begin{align}
&\frac{d}{dt}q_0(t) = -p_0\kappa q_0(t) +p_1 \kappa q_1(t) - q_0(t),\\
&\frac{d}{dt} q_1(t) =  -p_1\kappa q_1(t) + p_0\kappa q_0(t) + q_0(t),
\end{align}
\end{subequations}
where $q_0(t) = \text{Tr}[R_0(t)], q_1(t) = \text{Tr}[R_1(t)]$. Note also that $q_0(0) = \text{Tr}[\rho_0] = 1, q_1(0) = 0$. Furthermore, since $q_0(t) + q_1(t) = 1$, these equations can easily be integrated to obtain
\begin{align*}
&q_0(t) = \frac{1}{1 + \kappa(p_0 + p_1)} \bigg(\kappa p_1 + (1 + \kappa p_0) e^{-[1 + \kappa (p_0 + p_1)]t}\bigg), \\
&q_1(t) = \frac{1}{1 + \kappa(p_0 + p_1)} \bigg(1 - e^{-[1 + \kappa (p_0 + p_1)]t}\bigg).
\end{align*}
We note that both $q_0(t) = \text{Tr}[R_0(t)]$ and $q_1(t) = \text{Tr}[R_1(t)]$ are independent of $\rho_0$. Next, we integrate Eq.~\ref{eq:consequence_of_master_eq} to obtain that
\begin{align*}
&R_0(t) = \rho_0 e^{- t} -\kappa p_0 \int_0^t R_0(s) e^{-(t - s)}ds + \kappa p_1 \int_0^t R_1(s) e^{- (t - s)}ds + \kappa \sum_{i = 1}^k  \int_0^t e^{-(t - s)}   \big(\mathcal{N}_i - \textnormal{id}\big) R_0(s) ds, \\
&R_1(t) = \int_0^t U R_0(s) U^\dagger ds - \kappa p_1 \int_0^t R_1(s) ds + \kappa p_0 \int_0^t R_0(s) ds +  \kappa \sum_{i = 1}^k \int_0^t \big(\mathcal{N}_i - \textnormal{id}\big) R_1(s) ds.
\end{align*}
We note from these equations that since $\norm{R_0(t)}_1 , \norm{R_1(t)}_1 \leq 1$, 
\begin{align*}
&\norm{\int_0^t U R_0(s) U^\dagger ds - (1 - e^{-t}) U \rho_0 U^\dagger}_1, \norm{R_1(t) - \int_0^t U R_0(s) U^\dagger ds }_1 \leq 2\kappa t( k +1) 
\end{align*}
and consequently it follows from the triangle inequality that
\[
\norm{R_1(t) - (1 - e^{-t}) U \rho_0 U^\dagger}_1 \leq O(k\kappa),
\]
Now, we consider the $k-$qubit channel $\mathcal{E}$ given by $\mathcal{E}(\rho_0) := R_1(t) / q_1(t)$ --- this channel is the effective channel applied on the data qubits on post-selecting the clock qubit to be in $\ket{1}$. Furthermore, for $\rho_0 \succeq 0$ with $\text{Tr}[\rho_0] = 1$, it follows from the above estimates that
\[
\norm{\mathcal{E}(\rho_0) - U \rho_0 U^\dagger }_1 \leq \abs{1 - \frac{1 - e^{-t}}{q_1(t)}} \norm{U \rho_0 U^\dagger}_1 + \frac{1}{q_1(t)}\norm{R_1(t) - (1 - e^{-t}) U \rho_0 U^\dagger}_1 \leq O(\kappa k),
\]
where we have used that $\abs{q_1(t) - (1 - e^{-t})} \leq O(\kappa)$ and $q_1(t) = \Theta(1)$. Thus, we conclude that $\norm{\mathcal{E} - \mathcal{U}}_{1 \to 1} \leq O(k \kappa)$, hence proving the lemma statement. \hfill\(\square \)

\begin{reptheorem}{theorem:dissipative_dyn}[Formal] If $\mathcal{N}$ is the dephasing or amplitude damping channel, then for $d \geq 2$, $\exists \kappa_\textnormal{th} > 0$ such that for $\kappa \leq \kappa_\textnormal{th}$ problem \ref{prob:1} with purely dissipative Lindbladians cannot be weakly simulated on a classical computer within a multiplicative error\footnote{A family of $n-$qubit quantum circuits is said to be weakly simulable within multiplicative error $c$ if a classical computer can be used to sample in $\textnormal{poly}(n)$ time within a probability distribution $p_\textnormal{cl}$ such that
\[
\frac{1}{c} p(x) \leq p_\textnormal{cl}(x) \leq c p(x) \ \forall \ x \in \{0, 1\}^n,
\]
where $p$ is the probability distribution at the output of the quantum circuit.} $\in (1, \sqrt{2})$ unless the polynomial hierarchy collapses to the third level.
\end{reptheorem}
\noindent\emph{Proof}: It is previously known that if a family of quantum circuits, with post-selection (on $\textnormal{poly}(n)$ qubits, where $n$ is the problem size), solves a post-BQP complete problem then there isnt a classical algorithm that can weakly simulate it within a multiplicative error $\in (1, \sqrt{2})$ \cite{bremner2011classical}. Furthermore, post-selection has also been utilized in the context of noisy quantum circuits to show that below the error detection threshold, the circuit is expected to be clasically intractable \cite{fujii2016noise, fujii2016computational}. The key idea behind this proof is similar --- we implement the unitary circuit performing fault-tolerant quantum computation dissipatively together with post-selection on the clock qubits and thus show the low-noise intractability of the dissipative master equation.

Restricting ourselves to $d = 2$, we again will implement a 1D fault-tolerant quantum comptuation on a 2D grid --- one row of 2D grid of qubits is considered to be the data qubits, and the qubits in the columns are used as clock qubits. We initialize all the clock qubits to $\ket{0}$, and since both amplitude damping and dephasing channel have $\ket{0}\bra{0}$ as their fixed point, they remain in $\ket{0}\bra{0}$ uptil the point they are used to perform a unitary on the data qubits. Note that since we are allowed to use dissipative Lindbladians, the RESTART gates needed for the fault-tolerant quantum circuit can be trivially implemented by using an amplitude damping channel.  As outlined in lemma \ref{lemma:deph_unit}, this unitary can be applied dissipatively with $L = \ket{1}\bra{0} \otimes U$. The amplitude damping noise and the dephasing noise on the clock and the data qubits translates to faults in the effective unitary being applied. Furthermore, the faults appearing in the different unitaries (which use different clock qubits) are independent of each other.

After every time-step, we shift the clock qubits to provide a fresh clock qubit for the next time-step. Note that the clock qubit used to perform the unitary is in a mixture of $\ket{0}\bra{0}$ and $\ket{1}\bra{1}$, and is not entangled with the data qubits. Since we can only use dissipation, we need to perform the SHIFT operation amongst the clock qubits dissipatively --- furthermore, we need to perform this shift without additional clock qubits. As in the unitary case, we will perform this SHIFT operation with two layers of SWAP . To perform a SWAP operation dissipatively on two qubits, we use $L_1 = \ket{0, 1}\bra{1, 0}, L_2 = \ket{1, 0}\bra{0, 1} $. We note that with this master equation with $\mathcal{N}$ being dephasing noise, in time $\tau$,
\begin{subequations}\label{eq:diss_swap_deph}
\begin{align}
&\ket{0, 0}\bra{0, 0} \to \ket{0, 0}\bra{0, 0} , \ket{0, 1}\bra{0, 1} \to \frac{1}{2}\big(1 + e^{-2\tau}\big)\ket{0, 1}\bra{0, 1} + \frac{1}{2}\big(1 - e^{-2\tau}\big)  \ket{1, 0}\bra{1, 0}, \\
&\ket{1, 1}\bra{1, 1}\to \ket{1, 1}\bra{1, 1}, \ket{1, 0}\bra{1, 0} \to \frac{1}{2}\big(1 + e^{-2\tau}\big)\ket{1, 0}\bra{1, 0} + \frac{1}{2}\big(1 - e^{-2\tau}\big)  \ket{0, 1}\bra{0, 1}.
\end{align}
\end{subequations}
Furthermore, if $\mathcal{N}$ be amplitude damping noise, in time $\tau$, it follows that
\begin{subequations}\label{eq:diss_swap_amp_damp}
\begin{align}
&\ket{0, 0}\bra{0, 0} \to \ket{0, 0}\bra{0, 0}, \\
&\ket{0, 1}\bra{0, 1} \to (1 - e^{-\kappa \tau}) \ket{0, 0}\bra{0, 0} + \frac{e^{-\kappa \tau }}{2} \bigg((1 + e^{-2\tau}) \ket{0, 1}\bra{0, 1} + (1 - e^{-2\tau}) \ket{1, 0}\bra{1, 0}\bigg), \\
&\ket{1, 0}\bra{1, 0} \to (1 - e^{-\kappa \tau }) \ket{0, 0}\bra{0, 0} + \frac{e^{-\kappa \tau }}{2} \bigg((1 + e^{-2\tau}) \ket{1, 0}\bra{1, 0} + (1 - e^{-2\tau}) \ket{0, 1}\bra{0, 1}\bigg), \\
&\ket{1, 1}\bra{1, 1} \to e^{-2\kappa \tau} \ket{0, 0}\bra{0, 0} + 2e^{-\kappa \tau }\big(1 - e^{-\kappa \tau }\big)\big(\ket{1, 0}\bra{1, 0} + \ket{0, 1}\bra{0, 1}\big) + (1 -e^{-\kappa \tau })^2 \ket{1, 1}\bra{1, 1}.
\end{align}
\end{subequations}
It is important to note that in both of these cases, in a finite amount of time ($\tau = \Theta(1)$), the SHIFT operation does not succeed perfectly. However, it follows from Eqs.~\ref{eq:diss_swap_deph} and \ref{eq:diss_swap_amp_damp}, that it succeeds under postselection of the qubits that are in $\ket{1}$ in the following sense --- consider a set of $n$ qubits which are in a computational basis $\ket{b} = \ket{b_1, b_2 \dots b_n}$, and let an ideal SHIFT operation produce another computational basis $\ket{b'} = \ket{b_1', b_2' \dots b_n'}$. If $\rho$ is the state obtained on applying the dissipative SHIFT on $\ket{b}\bra{b}$, then $\rho$ conditioned on the qubits in $\ket{b'}$ that are in 1 is identical to $\ket{b'}\bra{b'}$ i.e.
\[
\bigg(\bigotimes_{i | b_i' = 1} \ket{1}_i\bra{1}\bigg) \rho \propto \ket{b'}\bra{b'}.
\]
With this shift operation, it is immediately clear that the purely dissipative master equation is post-BQP complete since applying any unitary dissipatively flips the clock qubit from $\ket{0}$ to $\ket{1}$ and then it is dissipatively shifted to replace the clock qubits. Postselecting on all the clock qubits to be in $\ket{1}$, we thus obtain the result of applying the fault tolerant circuit on the initial state, which completes the proof of this lemma. \hfill\(\square\)} 
\edit{\subsubsection{Impact of Lamb shifts and finite-temperature of the bath}
In our problem definition, we assumed the ability to implement an ideal dissipative master equation --- while from a theoretical standpoint, this could be a mathematical problem to be studied, in actual physical system the dissipative term can be accompanied by a coherent lamb shift, which adds a Hamiltonian term to the master equation. The proof of theorem 3 as described above does not account for this Hamiltonian term --- here, we show that despite this Lamb shift, the theorem \ref{theorem:dissipative_dyn} holds.

\emph{Master equation}: We first review the standard microscopic derivation of a master equation \cite{breuer2002theory, dann2018time} and show the corrections that arise to the model assumed in the previous subsection when the lamb shift and the non-zero environment temperatures are taken into account. Consider a system interacting with an environment through a possibly time-dependent jump operator $L(t)$. We assume a system-environment model of the form
\[
H(t) = L(t)\otimes B^\dagger(t) + L^\dagger(t) \otimes B(t).
\]
where $B(t)$ is an environment operator written in the interaction picture with respect to the Hamiltonian of the environment. We further assume that the jump operator is of the form $L(t) = \tilde{L}(t) e^{-i\varphi(t)}$, where $\tilde{L}(t)$ is a slowly varying operator on the time-scales of the relaxation times of the bath, and $\varphi(t)$ is a phase which can be possibly rapidly varying. We further assume that $\varphi'(t) = \omega_s(t) \geq 0$ --- this phase models the possibly time-dependent resonant frequencies at which the system oscillates. As is standard in the microscopic derivation of the master equation, we assume that system-environment density matrix factorizes at all times i.e. $\rho(t) \approx \rho_S(t) \otimes \rho_E$, where $\rho_S(t)$ is the system state at time $t$ and $\rho_E$ is the state of the environment. Then, we obtain that
\[
\frac{d}{dt}\rho_S(t) \approx -\int_0^t \text{Tr}_E\bigg([H(t), [H(t - s), \rho_S(t - s) \otimes \rho_E]]\bigg) ds,
\]
where we have made the assumption that $\text{Tr}(\rho_E B(t)) = 0$ on the state of the environment. Furthermore, also assuming that $\text{Tr}(\rho_E B^2(t)) = 0$ and making the Markovian approximation (i.e.~replacing $\rho_S(t - s)$ with $\rho_S(t)$ in the above equation), we obtain that
\begin{align*}
\frac{d}{dt}\rho_S(t) \approx -\sum_{\alpha \in \{+, -\}} \int_0^t \bigg[\Gamma_\alpha(-s)& \bigg(L_\alpha(t) L_\alpha^\dagger(t - s) \rho_S(t) - L_\alpha^\dagger(t - s) \rho_S(t) L_\alpha(t)\bigg)+\nonumber\\
& \Gamma_\alpha(s) \bigg(\rho_S(t) L_\alpha(t -s) L_\alpha^\dagger(t) - L_\alpha^\dagger(t) \rho_S(t) L_\alpha(t - s)\bigg) \bigg]ds
\end{align*}
where $L_+(t) = L(t), L_-(t) = L^\dagger(t)$ and, under the assumption that the environment is time-homogeneous, $\Gamma_-(s) = \Gamma_-^*(-s) = \text{Tr}(B^\dagger(t - s) B(t) \rho_E), \Gamma_+(s) = \Gamma_+^*(-s) = \text{Tr}(B(t - s) B^\dagger(t)\rho_E)$. We next perform the approximations
\[
L(t - s) = \tilde{L}(t - s) e^{-i\varphi(t - s)} \approx \tilde{L}(t) e^{-i(\varphi(t) - s \omega_S(t))} = L(t) e^{i s \omega_S(t)},
\]
and then we obtain that
\begin{subequations}\label{eq:microscopic_diss_meq}
\begin{align}
\frac{d}{dt}\rho_S(t) \approx \sum_{\alpha \in \{-, +\}} \kappa_\alpha(t) \bigg( L_\alpha(t) \rho_S(t) L_\alpha^\dagger(t) - \frac{1}{2}\{L_\alpha^\dagger(t) L_\alpha(t), \rho_S(t)\}\bigg) - i[\Delta_\alpha(t) L_\alpha^\dagger(t) L_\alpha(t), \rho_S(t)],
\end{align}
where
\begin{align}
\kappa_\pm(t) = 2\text{Re}\bigg(\int_0^t \Gamma_\pm(s) e^{i\omega_S(t) s} ds\bigg), \ \Delta_\pm(t) = \text{Im}\bigg(\int_0^t \Gamma_\pm(s) e^{i\omega_S(t) s} ds\bigg).
\end{align}
\end{subequations}
Compared to the master equation we used previously, which just had jump operator $L(t)$, Eq.~\ref{eq:microscopic_diss_meq} has two corrections --- first is another purely dissipative lindbladian with jump operator $L^\dagger(t)$. This arises due to the fact that an environment which is not unexcited can re-excite the system --- mathematically, this is a consequence of $\Gamma_-(s) \neq 0$, which is typical if the environment is at a finite-temperature as opposed to its vaccum state. The second difference is a Hamiltonian term with Hamiltonian $\Delta_+(t) L^\dagger(t) L(t) + \Delta_-(t) L(t) L^\dagger(t)$, which is the Lamb shift. The possibly time-dependent decay rates, $\kappa_\pm(t)$, are determined entirely by the spectral properties of the bath and the system resonant frequencies. For physically reasonable systems, we expect these rates to not be too large or too small --- we formalize this by demanding that there are $\kappa_\text{ub}, \kappa_\text{lb} >0$ such that
\begin{align}\label{eq:bounds_assumption_kappa}
(t_2 - t_1)\kappa_\text{lb} \leq\int_{t_1}^{t_2} \kappa_\pm(s) ds \leq(t_2 - t_1)  \kappa_\text{ub}.
\end{align}
This condition can be interpreted as demanding upper and lower bounds on the cumulative decay in a time-interval.

\emph{Analysis of the construction of theorem 3}: Here, we show that even if we use Eq.~\ref{eq:microscopic_diss_meq}, then the construction used in theorem 3 works. There are two parts to this --- first we show that lemma \ref{lemma:deph_unit} still holds, and then we show that the dissipative shift operation upon postselection also works. Consider first lemma \ref{lemma:deph_unit} --- here, we are using $L =   \ket{1}\bra{0} \otimes U$. As in lemma \ref{lemma:deph_unit}, we can assume that $R(t) = \ket{0}\bra{0} \otimes R_0(t) + \ket{1}\bra{1} \otimes R_1(t)$, and then we obtain that
\begin{subequations}\label{eq:diss_unitary_lamb}
\begin{align}
&\frac{d}{dt} R_0(t) = -p_0 \kappa R_0(t) + p_1 \kappa R_1(t) + \kappa \sum_{i = 1}^k \big(\mathcal{N}_i - \text{id}\big) R_0(t) + \kappa_-(t) U^\dagger R_1(t) U - \kappa_+(t) R_0(t), \\
&\frac{d}{dt} R_1(t) = -p_1 \kappa R_1(t) + p_0 \kappa R_0(t) + \kappa \sum_{i = 1}^k \big(\mathcal{N}_i - \text{id}\big) R_1(t) + \kappa_+(t) U R_0(t) U^\dagger - \kappa_-(t) R_1(t).
\end{align}
\end{subequations}
We note that, importantly, the Lamb shifts in the master equation do not impact the dynamics of $R(t)$ --- this is a consequence of the jump operator being a unitary on the data qubits, which results in the lamb shift Hamiltonian to be identity on them.

To analyze this system of equations, it is convenient to define the stochastic matrix $A(t)$ which is governed by the differential equation
\[
\frac{d}{dt} A(t) = \begin{bmatrix} -\kappa_+(t)  \text{id} & \kappa_-(t) \mathcal{U}^\dagger \\
\kappa_+(t) \mathcal{U} & -\kappa_-(t) \text{id}
\end{bmatrix} A(t),
\]
with $A(0) = I$. The $A(t)$ that solves this equation can be seen to be of the form
\[
A(t, s) = \begin{bmatrix} \big(1 - \alpha_0(t, s) \big) \text{id} & \alpha_1(t, s) \mathcal{U}^\dagger \\
\alpha_0(t, s) \mathcal{U} & \big(1 - \alpha_1(t, s)\big) \text{id}
\end{bmatrix}
\]
where $\mathcal{U}(\rho) = U\rho U^\dagger$ and $\alpha_0, \alpha_1$ are determined by the differential equations
\begin{align*}
\frac{d}{dt}\alpha_0(t, s) = \kappa_+(t) - \big(\kappa_+(t) + \kappa_-(t)\big)\alpha_0(t, s), \\ 
\frac{d}{dt}\alpha_1(t, s) = \kappa_-(t) - \big(\kappa_+(t) + \kappa_-(t)\big) \alpha_1(t, s),
\end{align*}
with boundary conditions $\alpha_0(s, s) = \alpha_1(s, s) = 0$. Eq.~\ref{eq:diss_unitary_lamb} can then be integrated to obtain
\[
\begin{bmatrix}
R_0(t) \\
R_1(t)
\end{bmatrix} = A(t, 0) \begin{bmatrix}
\rho_0 \\
0
\end{bmatrix} - \kappa \int_0^t A(t, s) \bigg(\begin{bmatrix} p_0 & -p_1 \\
-p_0 & p_1  
\end{bmatrix}\otimes \textnormal{id}\bigg)
\begin{bmatrix}
R_0(s) \\
R_1(s)
\end{bmatrix} ds+ \kappa \int_0^t A(t, s) \sum_{i = 1}^k \big( \mathcal{N}_i - \textnormal{id}\big)\begin{bmatrix}
R_0(s) \\
R_1(s)
\end{bmatrix} ds.
\]
Now,
\begin{align}\label{eq:lamb_shift_new}
\norm{R_1(t) - \alpha_0(t, 0) U \rho_0 U^\dagger}_1 \leq \norm{\begin{bmatrix}
R_0(t) \\
R_1(t)
\end{bmatrix}  - A(t, 0) \begin{bmatrix}
\rho_0 \\
0
\end{bmatrix}}_1 \leq 2t \kappa (k + 1).
\end{align}
Next we analyze the probability of obtaining the clock qubit in $\ket{1}$ --- let $q_0(t) = \text{Tr}[R_0(t)], q_1(t) = \text{Tr}[R_1(t)]$, we then obtain
\begin{align}\label{eq:diss_prob_dynamics}
&\frac{d}{dt}\begin{bmatrix}
q_0(t) \\
q_1(t)
\end{bmatrix} = \begin{bmatrix}
-\kappa_+(t) & \kappa_-(t) \\
\kappa_+(t) & -\kappa_-(t)
\end{bmatrix}
\begin{bmatrix}
q_0(t) \\
q_1(t)
\end{bmatrix} + \kappa
\begin{bmatrix}
p_0 & -p_1 \\
-p_0 & p_1
\end{bmatrix}
\begin{bmatrix}
q_0(t) \\
q_1(t)
\end{bmatrix} 
\end{align}
Noting that $q_0(0) = 1, q_1(0) = 0$, this system of equations can again be integrated to obtain
\[
\begin{bmatrix}
q_0(t) \\
q_1(t)
\end{bmatrix} =\begin{bmatrix}
1 - \alpha_0(t, 0) & \alpha_1(t, 0) \\
\alpha_0(t, 0) & 1 - \alpha_1(t, 0)
\end{bmatrix}
\begin{bmatrix}
1 \\
0
\end{bmatrix} +\kappa \int_0^t \begin{bmatrix}
1 - \alpha_0(t, s) & \alpha_1(t, s) \\
\alpha_0(t, s) & 1 - \alpha_1(t, s)
\end{bmatrix}\begin{bmatrix}
p_0 & -p_1 \\
-p_0 & p_1
\end{bmatrix}
\begin{bmatrix}
q_0(s) \\
q_1(s)
\end{bmatrix} ds,
\]
from which it follows that
\begin{align}\label{eq:prob_error_bound}
\abs{q_1(t) - \alpha_0(t, 0)} \leq 2\kappa t.
\end{align}
 Furthermore, since $q_0(t) + q_1(t) = 1$ we can also integrate Eq.~\ref{eq:diss_prob_dynamics} to obtain that
\begin{align}\label{eq:prob_lb}
q_1(t) = \int_0^t \big(p_0 \kappa + \kappa_+(t)\big) e^{-\int_s^t (\kappa + \kappa_+(s') + \kappa_-(s')) ds'} ds \implies q_1(t) \geq \kappa_\text{lb}e^{-t(2\kappa_\text{ub} + \kappa)},
\end{align}
where we have used Eq.~\ref{eq:bounds_assumption_kappa} for the lower bound. 

Finally, we bound the error between the post-selected quantum channel ($\mathcal{E}(\rho_0) = R_1(t) / q_1(t)$) and applying the unitary $U$. For $t = \Theta(1)$ (i.e.~some constant gate time) and from Eqs.~\ref{eq:lamb_shift_new}, \ref{eq:prob_error_bound} and \ref{eq:prob_lb}, we obtain that
\[
\norm{\mathcal{E}(\rho_0) - U \rho_0 U^\dagger}_1 \leq \frac{1}{q_1(t)} \norm{R_1(t) - \alpha_0(t, 0) U\rho_0 U^\dagger}_1 + \frac{1}{q_1(t)} \abs{\alpha_0(t,0) - q_1(t)} \leq O(k\kappa),
\]
which reproduces the conclusion of lemma \ref{lemma:deph_unit}.

Finally, to see that the shift operation on the clock qubits, upon post-selection for the outcome $\ket{1}$, we note first that the jump operators used in the SWAP operations involved in the shift operation are $\ket{1, 0}\bra{0, 1}$ and $\ket{0, 1}\bra{1, 0}$ --- these are already hermitian conjugates of each other and consequently the addition of an additional re-excitation term to the master equation does not impact the performance of the SWAP operation. Next, to see that the lamb shift also does not impact the SWAP operation, we note that during the SWAP operation, the state of the clock qubits being swapped remains in a state $\rho \in \text{span}\big\{\ket{0, 0}\bra{0, 0}, \ket{1, 1}\bra{1, 1}, \ket{1, 0}\bra{1, 0}, \ket{0, 1}\bra{0, 1}\big\}$. The lamb shift due to the jump operators $\ket{1, 0}\bra{0, 1}, \ket{0, 1}\bra{1, 0}$ is a Hamiltonian $H$$\in \text{span}\big\{\ket{1, 0}\bra{1, 0}, \ket{0, 1}\bra{0, 1}\big\}$. Since any such $\rho$ and $H$ commute, we conclude that the Lamb shift does not impact the SWAP operation.}
\bibliographystyle{unsrt}
\bibliography{references.bib}
\end{document}